\documentclass[11pt]{article}

\usepackage[preprint]{acl}

\usepackage{times}
\usepackage{latexsym}

\usepackage[T1]{fontenc}

\usepackage[utf8]{inputenc}

\usepackage{microtype}

\usepackage{inconsolata}

\usepackage{graphicx}
\usepackage{booktabs}
\usepackage{amsfonts}
\usepackage{amsmath}
\usepackage{graphicx}
\usepackage{subcaption}
\usepackage{tcolorbox}

\usepackage[shortlabels]{enumitem}

\newcommand{\name}[0]{AttentionRetriever}

\title{\name{}: Attention Layers are Secretly Long Document Retrievers}

\author{
 \textbf{David Jiahao Fu\textsuperscript{1}},
 \textbf{Lam Thanh Do\textsuperscript{1}},
 \textbf{Jiayu Li\textsuperscript{1}},
 \textbf{Kevin Chen-Chuan Chang\textsuperscript{1}}
\\
 \textsuperscript{1}University of Illinois Urbana-Champaign
 \\
 \texttt{\{jiahaof4, lamdo, jiayul11, kcchang\}@illinois.edu}
}

\begin{document}
\maketitle

\begin{abstract}

Retrieval augmented generation (RAG) has been widely adopted to help Large Language Models (LLMs) to process tasks involving long documents. However, existing retrieval models are not designed for long document retrieval and fail to address several key challenges of long document retrieval, including context-awareness, causal dependence, and scope of retrieval. In this paper, we proposed \name{}, a novel long document retrieval model that leverages attention mechanism and entity-based retrieval to build context-aware embeddings for long document and determine the scope of retrieval. With extensive experiments, we found \name{} is able to outperform existing retrieval models on long document retrieval datasets by a large margin while remaining as efficient as dense retrieval models.

\end{abstract}

\section{Introduction}
Recent advancements in Large Language Models (LLMs) \cite{openai_gpt-4_2023, openai_gpt-5_2025, dubey_llama_2024, jiang_mistral_2023} have demonstrated strong capabilities in natural language understanding, but recent studies \cite{liu_lost_2024, maharana_evaluating_2024, lu_controlled_2024} have shown that LLMs still struggle to perform well on long document processing tasks due to the lost-in-the-middle problem and limited context window length, while the quadratic complexity of the attention mechanism makes processing long documents very expensive, especially for large models with hundreds of billions of parameters. In recent years, retrieval-augmented generation (RAG) \cite{lewis_retrieval-augmented_2020} techniques have been widely applied to address this issue by selecting relevant information from the document with a retrieval model, which improves performance by removing distracting information and decreases processing time by shortening the input length. 

In the RAG pipeline, the retrieval model needs to perform \textbf{long document retrieval}, which requires the model to find a subset $D'$ from a user-provided long document $D$ such that $|D'| \ll |D|$ and $D'$ is sufficient and necessary to answer the input query $q$. However, existing retrieval models are not tailored for long document retrieval and overlook the following three types of dependencies in long documents:

\begin{itemize}
    \item \textbf{Contextual dependency.} Since long documents are generally coherent, context is often required to resolve issues like coreference and word ambiguity, which are crucial for determining the relevance of chunks. For example, in a document discussing Chicago, the author might use "the city" to refer to "Chicago", but this reference is clear only if the context is provided.  
    
    \item \textbf{Causal dependency.} The query may involve intermediate answers from the document that are needed to reach the final answer. For the same document about Chicago, an example query would be "What was the population of Chicago when the Great Fire happened?", where the intermediate answer "the Great Fire happened in 1871" is needed to find the chunk containing the final answer. 

    \item \textbf{Query dependency.} Text chunks providing background information, like the one containing "the Great Fire happened in 1871" in the previous example, are also important to answering the query and should be retrieved. However, these chunks might receive low similarity scores because they are not very relevant to the query, which asks about "the population of Chicago". Therefore, it is necessary to accurately decide the scope of retrieval in long document retrieval tasks.
\end{itemize}

Modeling the first two dependencies requires a more advanced retrieval model that can build context-aware representations and update the embeddings as additional contextual information is available. We found that the attention layers in transformer models perfectly match both requirements. Since attention layers calculate the representations of each token by aggregating information from other tokens, they are essentially cross-encoders that embed contextual information into the representation of each token, providing more abundant semantic information compared to existing embedding-based retrieval models. Furthermore, as the representations are propagated through layers, they are also dynamically adjusted based on contextual information gathered in previous layers to encode causal dependencies.

It is also intuitive to use attention layers as retrievers because the attention operations are essentially calculating similarity scores. In each attention layer of the transformer model, the attention score assigned to the $j$-th token by the $i$-th token is calculated as the weighted dot product between the key vector $k_j$ and the query vector $q_i$, which is identical to similarity calculation of embedding models. Moreover, attention computation is performed on two sets of embeddings $q$ and $k$, which allows the attention layers to perform a broader range of tasks other than semantic similarity search by adjusting the embeddings. 

However, training a transformer model for retrieval is very expensive. \citet{ye_infinite_2025} found that the last layer in the Qwen-2 model shows high retrieval accuracy without any additional training, implying the possibility of directly employing pretrained LLMs to estimate relevance with attention scores. However, their experiments were very limited and the findings might not generalize to other attention layers and other LLMs. More importantly, since pretrained LLMs suffer from accuracy and efficiency issues with long context, it is also crucial to find out whether attention scores can be efficiently calculated and still achieve high retrieval accuracy with long context. 

Therefore, we conducted careful analysis (Section \ref{sec:observation}) to verify if attention layers in pretrained LLMs can be effective and efficient training-free retrievers. Our analysis showed that \textbf{attention scores are more precise than outputs in collecting relevant information and suffer less from the lost-in-the-middle problem}, validating the effectiveness of using LLMs for retrieval tasks despite that they face various issues in long context tasks. Moreover, existing attention approximation methods for LLMs can be directly applied to this approach, making it possible to process long documents with arbitrary lengths. Furthermore, we found that employing pretrained LLMs with around 3 billion parameters as retrievers is already able to achieve impressive performance, which eliminates the need for using larger LLMs and improves efficiency. 

However, attention scoring alone is still insufficient to model the third dependency. Background information is still unlikely to receive high attention scores in layers with high retrieval accuracies because it is already embedded into the representations before these layers. To obtain a better estimate of the scope of retrieval, it is essential to additionally consider text chunks that are not immediately relevant but provide background information. Since each piece of background information typically focuses on one entity, and it should be included as part of the retrieval result only if the entity is relevant to the input query, we believe an entity graph structure could help to determine the retrieval scope precisely by connecting text chunks through entities and finding the entities relevant to the query during retrieval to discover hidden background information. In contrast to knowledge graphs, entity graphs are much easier and more efficient to construct, without the need to extract relationships between entities.  

\begin{figure*}[!htb]
    \centering
    \includegraphics[width=1\linewidth]{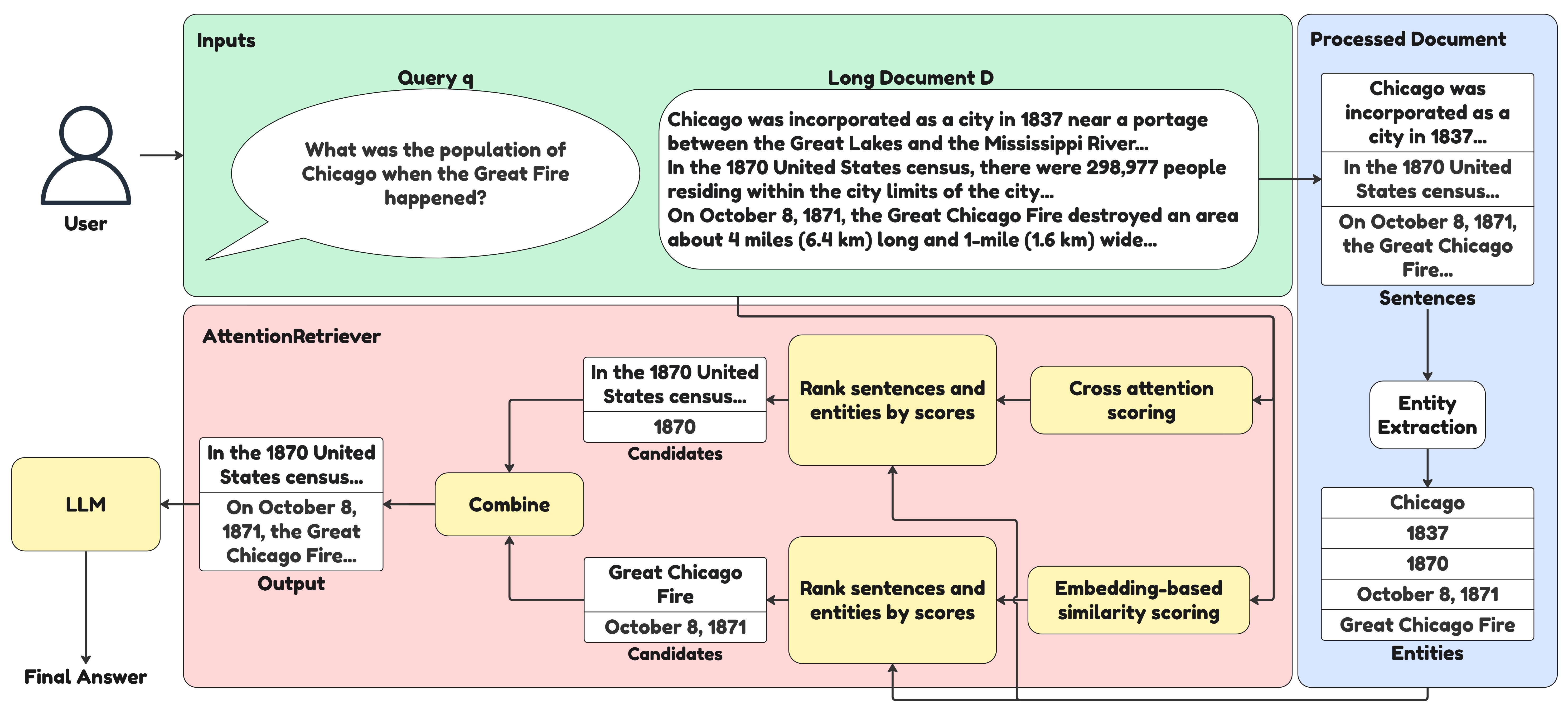}
    \caption{Overview of \name{}.}
    \label{fig:overview}
\end{figure*}

Based on these findings and analysis, we proposed \name{}, a novel retrieval model that builds context-aware embeddings with pretrained LLMs and decides the scope of retrieval through entity-based retrieval, as summarized in Figure \ref{fig:overview}. During retrieval, we leverage LLMs to process the long document and the query together, adopting the attention maps at layers that show high retrieval accuracies to estimate the relevance score of each text segment, which is combined with embedding-based similarity scoring for more precise retrieval. To decide the scope of retrieval, we find the desired entities by ranking them by the scores of sentences containing the entities. We eventually obtain the final output by collecting all text chunks that contain the highest ranked entities and sentences.

To better evaluate and compare the performance and efficiency of our proposed method and baselines on extremely long documents, we also constructed a new long document retrieval dataset consisting of different types of documents with an average length of over 100,000 words and various types of queries (Section \ref{sec:dataset}). To the best of our knowledge, this is the first retrieval dataset that features documents with lengths exceeding the context window length of most existing LLMs.

We evaluated \name{} on our constructed dataset, as well as six other single-document retrieval datasets and three multi-document retrieval datasets. \name{} outperforms state-of-the-art sparse and dense retrieval models by a large margin on single-document retrieval datasets while remaining as efficient as dense retrieval models with similar sizes. Moreover, \name{} also achieves competitive performance in multi-document retrieval datasets where contextual information is generally not needed, further demonstrating its effectiveness in long document retrieval tasks.

In summary, our contributions are as follows:
\begin{itemize}
    \item We propose \name{}, which leverages the attention mechanism and entity-based retrieval to perform context-aware long document retrieval and decide the scope of retrieval based on the query;
    \item We perform empirical analysis on the attention mechanism in pretrained LLMs, verifying the effectiveness of using attention maps for retrieval tasks and its capability of dynamically updating embeddings across layers;
    \item We construct a new long document retrieval dataset consisting of extremely long documents to compare the retrieval accuracy of our proposed method and various baselines.
\end{itemize}

\section{Related Works}
\subsection{Long Document Retrieval}

Although the concept of text retrieval has appeared for many years, the task of long document retrieval remains largely unexplored. Existing sparse and dense models, such as BM25 \cite{robertson_probabilistic_2009}, DPR \cite{karpukhin_dense_2020}, ANCE \cite{xiong_approximate_2021}, GTR \cite{ni_large_2022}, mGTE \cite{zhang_mgte_2024}, and Grit-LM \cite{muennighoff_generative_2025}, are mostly designed for open-domain retrieval, in which the models deal with a large corpus of independent documents instead of a long document and can process each document separately because other ones are likely to be irrelevant. Researchers have attempted to incorporate context-awareness into retrieval models \cite{morris_contextual_2024, gunther_late_2024, conti_context_2025}, but they are still designed for open-domain retrieval. In recent years, SPScanner \cite{cao_single-pass_2025} and MC-Indexing \cite{dong_mc-indexing_2024} have been proposed to address the problem of long document retrieval. However, they still failed to address the challenges of causal and query dependencies. 

\subsection{Context Window Length Extension}

LLMs fail to process inputs exceeding its context window length because they lack training on out-of-distribution (OOD) indices. To tackle this problem, prior works \cite{ding_longrope_2024,shang_longrope2_2025,an_training-free_2024,jin_llm_2024,xu_extending_2025,liu_reattention_2025} proposed to map OOD indices to in-distribution indices to avoid the problem. To reduce the computational cost when processing long contexts, researchers also proposed methods to approximate the full attention map by dividing the context into text segments and performing attention operations only on relevant text segments \cite{xiao_infllm_2024, fountas_human-like_2024, willette_training_2025}. 

\subsection{Attention Mechanism Interpretation}

Despite the success of modern LLMs, it remains unclear how the attention mechanism \cite{bahdanau_neural_2015} in LLMs contribute to its success. \citet{vig_analyzing_2019} showed that different attention heads are assigned to different tasks in GPT-2, while \citet{sun_transformer_2024} also found that the middle transformer layers have similar functionalities and are responsible for different tasks. \citet{ye_infinite_2025} conducted analysis on the final layer of Qwen-2 model and concluded that high attention scores are given to tokens relevant to the query. However, no existing work has explored the potential of employing attention layers in LLMs for retrieval. 

\section{Observations}
\label{sec:observation}

In order to verify whether the attention layers of different LLMs can accurately find relevant chunks for different types of queries, we conducted a detailed empirical analysis to measure the effectiveness of each attention layer for retrieval and the shift of attention patterns across layers. We selected LLaMA-3.2 3B \cite{dubey_llama_2024}, Qwen-2.5 3B \cite{yang_qwen25_2024}, and Mistral 7B \cite{jiang_mistral_2023} as representatives of LLMs, and analyzed the cross attention between the query and the document at each attention layer when the document and the query are processed by the LLMs. To compare the attention patterns of different types of queries, we utilized the training set of MuSiQue dataset \cite{trivedi_musique_2022}, which contains different types of multi-hop questions and can be decomposed into simpler subqueries. 

In each attention layer, we calculate its retrieval accuracy by finding the paragraphs the model focuses on through cross attention scores. Specifically, we estimate the relevance of each paragraph by the maximum cross attention scores assigned by the query to tokens from the paragraph, averaged over all attention heads. Formally, given the attention map $A\in\mathbb{R}^{H\times T_d \times T_q}$, the paragraph score $score_p$ is calculated as:
$$score_p=\max_{s_l\leq t\leq s_r, 1\leq t_q\leq T_q}\left(\frac 1 H \sum_{h=1}^H A_{h, t, t_q}\right)$$
where $H$ is the number of attention heads, $T_d$ is the length of the document, $T_q$ is the length of the query, and the paragraph $p$ spans over tokens $p_l, \dots, p_r$. We then rank the paragraphs by their corresponding scores to determine the paragraphs being focused on and compare with the gold paragraphs for each subquery of the questions to calculate the retrieval accuracy for the subquery. 

\begin{figure}[t!]
    \centering
    \begin{subfigure}[t]{\linewidth}
        \includegraphics[width=\linewidth]{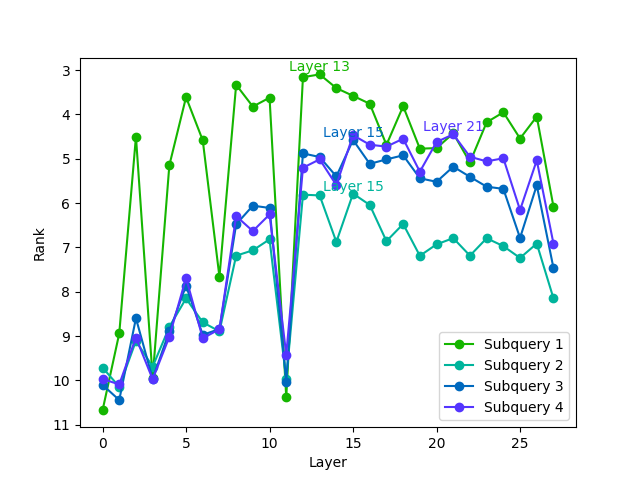}
        \caption{Average ranks of the gold paragraph for each subquery over all queries in the dataset. The layer achieving the highest average rank for each subquery is marked at the top of each line. The subqueries are ordered by dependencies, where subquery 1 does not depend on any other subqueries while subquery 4 could depend on any other subqueries.}    
        \label{fig:analysis-a}
    \end{subfigure}
    \begin{subfigure}[t]{\linewidth}
        \includegraphics[width=\linewidth]{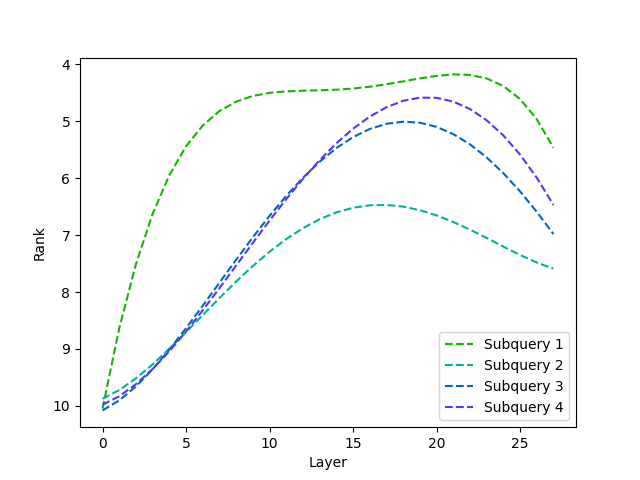}
        \caption{The quartic approximation of the average ranks.}
        \label{fig:analysis-b}
    \end{subfigure}
    \caption{Results of attention analysis.}
    \label{fig:analysis}
\end{figure}

Figure \ref{fig:analysis} presents the results of attention layers of \texttt{Llama-3.2-3B-Instruct} on all queries, while the results for other models and each specific type of queries can be found in Appendix \ref{appendix:analysis}. The top figure shows the average ranking of the gold paragraph for each subquery across the layers, while the bottom figure is the quartic approximation of the rankings. From Figure \ref{fig:analysis-a}, we found that \textbf{only certain attention layers can achieve high retrieval accuracy}, and they are mostly in the second half of the layers. By analyzing the approximation in Figure \ref{fig:analysis-b}, we also found that \textbf{pretrained LLMs shift focuses in different attention layers}. Earlier attention layers tend to focus on the independent subquery (subquery 1), while later layers rank the gold paragraphs for subqueries 2, 3, and 4 higher, which depends on subquery 1. This result validates the process of building context and causal dependencies through attention layers in LLMs.

\begin{figure}[t!]
    \centering
    \begin{subfigure}[t]{\linewidth}
        \includegraphics[width=\linewidth]{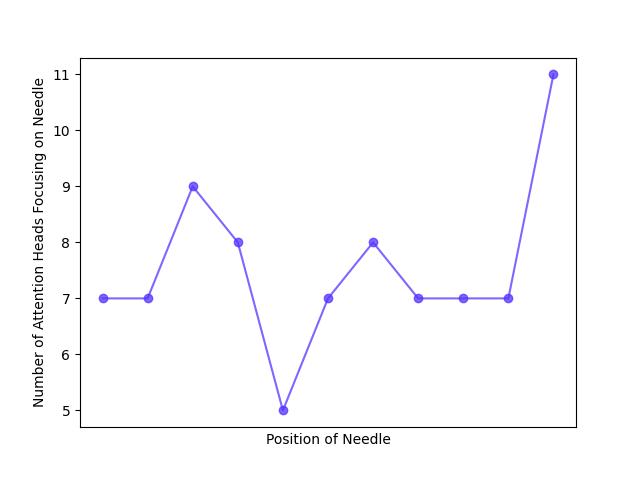}
        \caption{Retrieval accuracy for attention layers.}
        \label{fig:niah-a}
    \end{subfigure}
    \begin{subfigure}[t]{\linewidth}
        \includegraphics[width=\linewidth]{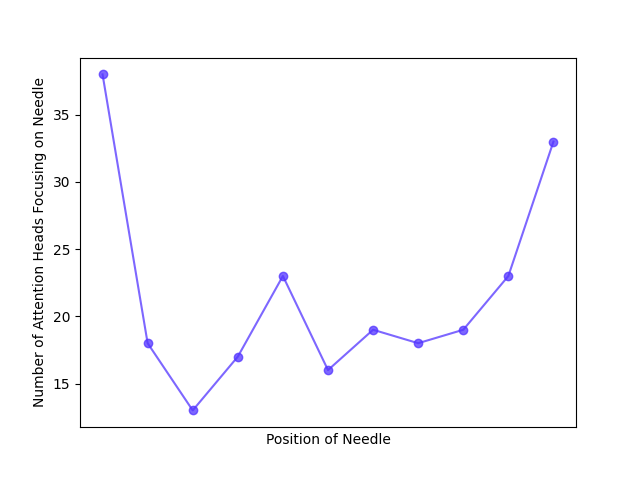}
        \caption{Retrieval accuracy for attention layers with Cascading KV Cache.}
        \label{fig:niah-b}
    \end{subfigure}
    \caption{Test results on the needle-in-a-haystack test.}
    \label{fig:niah}
\end{figure}

However, pretrained LLMs are known to struggle with processing long documents due to lost-in-the-middle problem and limited context window length. Therefore, we conducted additional experiments on the needle-in-a-haystack \cite{noauthor_gkamradtllmtest_needleinahaystack_nodate} test to verify if lost-in-the-middle also appears in attention layers and if existing context length extension methods can be applied to our pipeline. To determine whether the attention layers can find the needles, we calculated the number of attention heads where the highest attention score is assigned to the needle. We selected \texttt{Llama-3.2-3B-Instruct} as the base model and employed Cascading KV Cache \cite{willette_training_2025} as the context extension method, and tested on documents with approximately 100,000 tokens. 

Figure \ref{fig:niah} shows the results of our experiments. Figure \ref{fig:niah-a} reveals that the count does not decrease when the needle is in the middle of the document, indicating that \textbf{attention layers are less affected by the lost-in-the-middle problem}. Figure \ref{fig:niah-b} also demonstrates that the approximation method can be effective in our proposed pipeline because it can even find needles much more accurately than full attention. Based on these observations, we proposed our attention-based retrieval approach, which will be outlined in the next section.

\section{Method}
\subsection{Overview}

Our proposed retrieval model is summarized in Figure \ref{fig:overview}. We leverage a pretrained LLM to assign a score for each sentence in the document based on attention maps of the LLM (Subsection \ref{subsec:attn}). Meanwhile, we also employ a dense embedding model to calculate a separate score for each sentence by similarity between the sentence embedding the and query embedding to further improve retrieval accuracy (Subsection \ref{subsec:emb}). We then use these two sets of scores to find relevant entities and perform an entity-based retrieval to obtain the outputs (Subsection \ref{subsec:retrieval}).

\subsection{Attention for Sentence Scoring}
\label{subsec:attn}

The observations of Section \ref{sec:observation} suggest that the cross-attention scores between the query and the document in pretrained LLMs could provide an accurate estimate of query relevance. Based on this conclusion, we proposed to employ a pretrained LLM to estimate the relevance of each sentence by computing cross-attention scores between tokens corresponding to the query and tokens from the sentence. Since only certain layers achieve high retrieval accuracies, we only use the attention scores from layers that obtained smallest average ranking for at least one subquery (the layers labeled in Figure \ref{fig:analysis-a}) to remove the noise of other layers.

We followed the same procedure as Section \ref{sec:observation} to convert token-level attention scores into sentence scores to fully utilize our findings. After obtaining the raw attention map $A\in\mathbb{R}^{L\times H\times T_d \times T_q}$, where $L$ is the number of selected layers, $H$ is the number of attention heads, $T_d$ is the length of the document, and $T_q$ is the length of the query, we calculate the attention score for each sentence by taking the maximum attention score assigned to tokens in the sentence across all selected layers and all query tokens, averaged over attention heads. More formally, given a sentence $s$ spanning over tokens $s_l, \dots, s_r$, the attention score of the sentence $a_s$ is computed as follows:

$$a_s = \max_{1\leq l\leq L, s_l\leq t\leq s_r, 1\leq t_q\leq T_q}\left(\frac 1 H \sum_{h=1}^H A_{l, h, t, t_q}\right)$$

In Section \ref{sec:observation}, we also found that Cascading KV cache method \cite{willette_training_2025} can be seamlessly applied to our method, so we apply this method to handle inputs of that could exceed the pre-defined context window limit of our selected base model efficiently. In practice, since the extension method would bring additional overhead to the pipeline, we only apply it when the length of the document exceeds the context window limit of the base LLM. 

\subsection{Sentence Embedding for Multi-view Similarity Search}
\label{subsec:emb}

Attention-based similarity estimation provides a token-level measurement of the relevance to the query, which is in fact complementary to the traditional embedding-based sentence-level estimation. To enhance our retrieval process, we use sentence embedding to obtain a sentence-level similarity estimate as an additional view in similarity search to enrich the information obtained by attention-based search. We encode each sentence $s$ into its embedding $E_s = f(s)$ through an embedding model $f$, and use the same embedding model to obtain the query embedding query embedding $E_q = f(q)$. The sentence embedding score $e_s$ for each sentence $s$ is calculated as the cosine similarity between its embedding $E_s$ and query embedding $E_q$:

$$e_s = \frac{E_s \cdot E_q}{\lvert\lvert E_s\rvert\rvert \space \lvert\lvert E_q\rvert\rvert}$$

\subsection{Entity-based Retrieval}
\label{subsec:retrieval}

In long document retrieval, it is insufficient to only retrieve the most relevant text chunks because long documents are typically coherent and other chunks could provide additional background information about the query. Since chunks mentioning the entities relevant to the query can usually provide useful information, we find the most relevant entities and use them to find these indirectly relevant text chunks. 

Specifically, we use SpaCy \cite{honnibal_spacy_2020} to extract the entities in each sentence, and assign a relevance score to each entity based on the scores of the sentences it appears in to find the most relevant entities. Since relevant entities should only appear in relevant sentences, we calculate the relevance score of entities by the average relevance scores of the sentences.

Since we obtained both attention scores and embedding scores and the attention score and embedding score are not directly comparable, we perform retrieval on each score separately and combine the results with equal weights. With a \texttt{top\_k} value of $k$, we first retrieve $\lceil \frac k 2\rceil$ entities and sentences with the highest attention scores and $\lfloor \frac k 2\rfloor$ entities and sentences with the highest embedding scores. The union of these two sets of entities and sentences is the collection of all selected entities and sentences. For each selected sentence, we retrieve the paragraph it belongs to, while for each selected entity, we retrieve all paragraphs containing the entity.

\begin{table*}[!htb]
    \centering
    \tiny
    \resizebox{\textwidth}{!}{
        \begin{tabular}{l|ccccccc|c}
            \toprule
            & QASA & Qasper & RepLiQA & ConditionalQA & NaturalQuestions & LongBench-v2-Retrieval & Average \\
            Average Length & 4665.09 & 3442.26 & 970.50 & 1298.13 & 2548.97 & 106025.49 & \\
            \midrule
            \textbf{Sparse Models} & & & & & & & \\
            \quad BM25 & 0.3795 & 0.2304 & 0.4896 & 0.1988 & 0.3055 & 0.3126 & 0.3194 \\
            \midrule
            \textbf{Dense Models} & & & & & & & \\
            \quad DPR & 0.2181 & 0.0529 & 0.3602 & 0.1542 & 0.2981 & 0.1226 & 0.2010 \\
            \quad ANCE & 0.3724 & 0.2856 & 0.4686 & 0.1893 & 0.3818 & 0.3240 & 0.3370 \\
            \quad CDE & 0.1341 & 0.0962 & 0.2449 & 0.1547 & 0.2416 & 0.0485 & 0.1532 \\
            \quad GTR & 0.3977 & 0.2714 & 0.4851 & 0.2584 & 0.4110 & 0.3260 & 0.3583 \\
            \quad GTE-Qwen2 & 0.3785 & 0.2442 & 0.4785 & 0.2415 & 0.4131 & 0.3314 & 0.3479 \\
            \quad Qwen3 & 0.4414 & 0.2057 & 0.4846 & 0.2892 & 0.4091 & 0.3205 & 0.3584 \\
            \quad GritLM & 0.4394 & 0.3008 & 0.5141 & \underline{0.3258} & 0.4592 & 0.3398 & 0.3965 \\
            \midrule
            \textbf{Autoregressive Models} & & & & & & & \\
            \quad SPScanner & 0.4604 & 0.3712 & 0.6434 & 0.1354 & 0.4237 & \underline{0.4188} & 0.4088 \\
            \midrule
            \textbf{\name{} (Ours)} & & & & & & & \\
            \quad LLaMA-3.2 3B & \underline{0.5584} & \textbf{0.4618} & \underline{0.8339} & \textbf{0.3526} & \textbf{0.5998} & \textbf{0.4738} & \textbf{0.5467} \\
            \quad Qwen-2.5 3B & \textbf{0.5663} & \underline{0.4551} & \textbf{0.8422} & 0.3106 & \underline{0.5655} & 0.3215 & \underline{0.5102} \\
            \bottomrule
        \end{tabular}
    }
    \caption{Comparison of proposed method and baselines on single-document retrieval datasets, where best and second best results are marked in \textbf{bold} and \underline{underline}, respectively.}
    \label{table:retrieval-result-single}
\end{table*}

\section{Dataset Construction}
\label{sec:dataset}

Since the average length of documents in existing retrieval datasets are very limited, we collected a set of long documents from LongBench-v2 dataset \cite{bai_longbench_2025}, which features long documents from a variety of sources, to compare our proposed method with baselines on extremely long documents. We only selected documents with length labeled as \texttt{"medium"} or \texttt{"long"}, and sub-domain is one of \texttt{Financial}, \texttt{Academic}, \texttt{Governmental}, and \texttt{Legal} to obtain more well-structured documents. We collected a total of 35 documents from the dataset.

Through our analysis on prior retrieval and QA datasets, such as MuSiQue \cite{trivedi_musique_2022}, Qasper \cite{dasigi_dataset_2021}, and LongBench \cite{bai_longbench_2024}, we found that most questions in the datasets fall into the following four categories:

\begin{itemize}
    \item Single-hop: the question can be answered with a single piece of information.
    \item Comparison: the question requires comparison between two or more pieces of information from multiple paragraphs.
    \item Composition: the question composes multiple single-hop questions, and the LLM needs to answer each sub-question in sequence to arrive at the final answer.
    \item Summarization: the question requires summarizing information from multiple locations throughout the document.
\end{itemize}

Therefore, to ensure our dataset includes all these four types of queries, we manually generated a query for each of these four types of queries for each document we collected, and annotated the corresponding answer and relevant paragraphs. We denote our dataset as \textbf{LongBench-v2-Retrieval} for the rest of the paper. A comparison of our dataset and other long document retrieval datasets can be found in Appendix \ref{appendix:dataset}, which shows that the documents in our new dataset are significantly longer than other retrieval datasets and some of them even exceed the context window length of most LLMs \cite{dubey_llama_2024, jiang_mistral_2023, zhang_qwen3_2025}. 

\section{Experiments}
\subsection{Experimental Setup}

To comprehensively evaluate the effectiveness and efficiency of our proposed method, we conducted our experiment on both long document retrieval tasks and long document question answering (QA) tasks with the retrieval-augmented generation (RAG) setting.

\textbf{Baselines.} We compared \name{} with the following three types of existing retrieval models that can be applied to long document retrieval:

\begin{enumerate}[label=(\alph*)]
    \item Sparse retrieval model BM25 \cite{robertson_probabilistic_2009};
    \item Dense retrieval models, such as DPR \cite{karpukhin_dense_2020}, ANCE \cite{xiong_approximate_2021}, GTR \cite{ni_large_2022}, GTE-Qwen2-7B \cite{zhang_mgte_2024}, GritLM-7B \cite{muennighoff_generative_2025}, and Qwen3-7B \cite{zhang_qwen3_2025};
    \item Long document retrieval model SPScanner-1.3B \cite{cao_single-pass_2025};
\end{enumerate}

For evaluation on QA tasks, we used pretrained LLMs, such as LLaMA-3.1 8B \cite{dubey_llama_2024}, Mistral-7B \cite{jiang_mistral_2023}, Qwen-2.5 7B \cite{yang_qwen25_2024}, and GPT-5 mini \cite{openai_gpt-5_2025}, as our baselines to compare between RAG setting and direct generation.

\textbf{Datasets.} We conducted experiments on five long document retrieval benchmarks: LongBench-v2-Retrieval (Ours), QASA \cite{lee_qasa_2023}, Qasper \cite{dasigi_dataset_2021}, RepLiQA \cite{monteiro_repliqa_2024}, and DAPR \cite{wang_dapr_2024}. For DAPR, we only selected datasets where each query can be answered by a single document, which are ConditionalQA, MS MARCO, and Natural Questions. However, we decided not to use MS MARCO because the average paragraph count of documents in the datasets is very limited (as shown in Table \ref{table:dataset-stat}), and is not suitable for our task. We report the performance of our proposed method and the baselines on the test sets of these benchmarks. For LongBench-v2-Retrieval and RepLiQA, we used the full datasets because the dataset splits are not provided. 

We also used three multi-document retrieval benchmarks for further comparison, which are HotpotQA \cite{yang_hotpotqa_2018}, 2WikiMultihopQA \cite{ho_constructing_2020}, and MuSiQue \cite{trivedi_musique_2022}. We report the performance of our proposed method and the baselines on the validation sets of these benchmarks. 

For evaluation on the QA task, we utilized the LongBench \cite{bai_longbench_2024} dataset and selected three single-document QA subsets, Qasper, MultiFieldQA-en, and NarrativeQA. 

\textbf{Metrics.} To evaluate and compare retrieval and QA performance, we use \textbf{F-1} as the evaluation metric. For comparison of retrieval efficiency, we measure the average processing time for each sample in the datasets. For QA tasks, we report the number of input tokens passed into the QA model, as we cannot measure the processing time for closed-source models. 

\textbf{Implementation Details.} In our experiments, we use paragraphs as basic retrieval units. For baselines that cannot process long paragraphs, we split each paragraph into text chunks that can fit into the context window and take the maximum similarity of the text chunks with the query as the similarity of the paragraph. For \name{}, we used \texttt{Llama-3.2-3B-Instruct} and \texttt{Qwen/Qwen2.5-3B-Instruct} as our base LLMs to compute attention scores. For \texttt{top\_k} choices, we used 1, 2, 3, and 5 because the average number of evidences of the datasets we used are below 5 (as shown in Table \ref{table:dataset-stat}). For QA tasks, we also used vLLM \cite{kwon_efficient_2023} for efficient generation. We performed all our experiments on a single NVIDIA A40 GPU. 

\subsection{Main Results} Table \ref{table:retrieval-result-single} and \ref{table:retrieval-result-multi} present the retrieval performance of baselines and \name{} on single-document and multi-document retrieval datasets, respectively. For simplicity, we only included the results with \texttt{top\_k} value of 3, while the results of other \texttt{top\_k} values can be found in Appendix \ref{appendix:retrieval}. Our proposed method significantly outperforms the baselines across all single-document retrieval datasets, while achieving a similar performance with baselines on multi-document datasets, which is not the primary target of our method. However, we also noticed a significant performance drop of \name{}-Qwen on LongBench-v2-Retrieval dataset compared to other dataset, which could indicate that Qwen-2.5 model does not work well with the context extension method. Table \ref{table:retrieval-result-efficiency} compares the retrieval efficiency of \name{} and the baselines. Our proposed method is as efficient as large dense embedding models like GTE, Qwen3, and GritLM, demonstrating the efficiency of \name{}. 

The results for the QA tasks can be found in Appendix \ref{appendix:qa}. Our proposed method achieves comparable performance as direct generation with the LLM while significantly reducing the number of input tokens, while also outperforming the long document retrieval baseline SPScanner. Notably, we found that all RAG methods fail to perform well on NarrativeQA dataset, which might indicate that RAG methods do not work well on novels since novels are typically less well-structured than other types of long documents and retrieval cannot collect all pieces of information required to answer the question. 

\begin{table}
    \centering
    \tiny
    \resizebox{\columnwidth}{!}{
        \begin{tabular}{l|ccc|c}
            \toprule
            & HotpotQA & 2WikiMultihopQA & MuSiQue & Average \\
            Average Length & 887.05 & 549.58 & 8089.59 & \\
            \midrule
            \textbf{Sparse Models} & & & \\
            \quad BM25 & 0.5695 & 0.5454 & 0.3412 & 0.4854 \\
            \midrule
            \textbf{Dense Models} & & & \\
            \quad DPR & 0.5020 & 0.5389 & 0.3680 & 0.4696 \\
            \quad ANCE & 0.5547 & 0.6603 & 0.4285 & 0.5478 \\
            \quad CDE & 0.3932 & 0.2637 & 0.2320 & 0.2963 \\
            \quad GTR & 0.6109 & 0.6393 & 0.4591 & 0.5698 \\
            \quad GTE-Qwen2 & 0.6707 & \underline{0.664} & \underline{0.5328} & \underline{0.6225} \\
            \quad Qwen3 & 0.6721 & 0.6533 & 0.522 & 0.6158 \\
            \quad GritLM & \textbf{0.7096} & \textbf{0.6802} & \textbf{0.5484} & \textbf{0.6461} \\
            \midrule
            \textbf{Autoregressive Models} & & & \\
            \quad SPScanner & 0.6052 & 0.6268 & 0.3982 & 0.5434 \\
            \midrule
            \textbf{\name{} (Ours)} & & & \\
            \quad LLaMA-3.2 3B & \underline{0.7090} & 0.6495 & 0.5084 & 0.6223 \\
            \quad Qwen-2.5 3B & 0.7037 & 0.6355 & 0.5062 & 0.6151 \\
            \bottomrule
        \end{tabular}
    }
    \caption{Comparison of proposed method and baselines on multi-document retrieval datasets, where best and second best results are marked in \textbf{bold} and \underline{underline}, respectively.}
    \label{table:retrieval-result-multi}
\end{table}

\section{Conclusion}
In this paper, we introduced \name{}, a training-free context-aware retrieval model that leverages the attention mechanism in LLMs and entity-based retrieval for accurate long document retrieval. In contrast to prior sparse and dense retrieval models, our approach considers contextual, causal, and query dependencies in the long document when evaluating the relevance of each text segment, enabling a more precise retrieval process. Our experiment results reveal that \name{} can achieve very strong performance on various long document retrieval benchmarks, while maintaining competitive performances on multi-document retrieval scenarios. 

\section*{Limitations}
We identified several major limitations of our work. Firstly, our proposed method requires a sufficiently large LLM (around 3 billion parameters) to perform well, making it less efficient than sparse and small dense models. Secondly, due to hardware constraints, we did not extend our analysis on attention to larger LLMs, which could show different attention patterns. Lastly, due to the scarcity of well-formatted long documents and the difficulty of manual labeling, the size of the retrieval dataset we constructed is limited and can be sensitive to outliers, causing the evaluation results to be inaccurate. 

\bibliography{custom}

@article{honnibal_spacy_2020,
	title = {{spaCy}: {Industrial}-strength {Natural} {Language} {Processing} in {Python}},
	doi = {10.5281/zenodo.1212303},
	author = {Honnibal, Matthew and Montani, Ines and Van Landeghem, Sofie and Boyd, Adriane},
	year = {2020},
}

@article{robertson_probabilistic_2009,
	title = {The {Probabilistic} {Relevance} {Framework}: {BM25} and {Beyond}},
	volume = {3},
	url = {https://doi.org/10.1561/1500000019},
	doi = {10.1561/1500000019},
	number = {4},
	journal = {Found. Trends Inf. Retr.},
	author = {Robertson, Stephen E. and Zaragoza, Hugo},
	year = {2009},
	pages = {333--389},
}

@inproceedings{ding_longrope_2024,
	title = {{LongRoPE}: {Extending} {LLM} {Context} {Window} {Beyond} 2 {Million} {Tokens}},
	url = {https://openreview.net/forum?id=ONOtpXLqqw},
	booktitle = {Forty-first {International} {Conference} on {Machine} {Learning}, {ICML} 2024, {Vienna}, {Austria}, {July} 21-27, 2024},
	publisher = {OpenReview.net},
	author = {Ding, Yiran and Zhang, Li Lyna and Zhang, Chengruidong and Xu, Yuanyuan and Shang, Ning and Xu, Jiahang and Yang, Fan and Yang, Mao},
	year = {2024},
}

@inproceedings{maharana_evaluating_2024,
	title = {Evaluating {Very} {Long}-{Term} {Conversational} {Memory} of {LLM} {Agents}},
	url = {https://doi.org/10.18653/v1/2024.acl-long.747},
	doi = {10.18653/V1/2024.ACL-LONG.747},
	booktitle = {Proceedings of the 62nd {Annual} {Meeting} of the {Association} for {Computational} {Linguistics} ({Volume} 1: {Long} {Papers}), {ACL} 2024, {Bangkok}, {Thailand}, {August} 11-16, 2024},
	publisher = {Association for Computational Linguistics},
	author = {Maharana, Adyasha and Lee, Dong-Ho and Tulyakov, Sergey and Bansal, Mohit and Barbieri, Francesco and Fang, Yuwei},
	editor = {Ku, Lun-Wei and Martins, Andre and Srikumar, Vivek},
	year = {2024},
	pages = {13851--13870},
}

@article{fountas_human-like_2024,
	title = {Human-like {Episodic} {Memory} for {Infinite} {Context} {LLMs}},
	volume = {abs/2407.09450},
	url = {https://doi.org/10.48550/arXiv.2407.09450},
	doi = {10.48550/ARXIV.2407.09450},
	journal = {CoRR},
	author = {Fountas, Zafeirios and Benfeghoul, Martin and Oomerjee, Adnan and Christopoulou, Fenia and Lampouras, Gerasimos and Bou-Ammar, Haitham and Wang, Jun},
	year = {2024},
	note = {arXiv: 2407.09450},
}

@article{xiao_infllm_2024,
	title = {{InfLLM}: {Unveiling} the {Intrinsic} {Capacity} of {LLMs} for {Understanding} {Extremely} {Long} {Sequences} with {Training}-{Free} {Memory}},
	volume = {abs/2402.04617},
	url = {https://doi.org/10.48550/arXiv.2402.04617},
	doi = {10.48550/ARXIV.2402.04617},
	journal = {CoRR},
	author = {Xiao, Chaojun and Zhang, Pengle and Han, Xu and Xiao, Guangxuan and Lin, Yankai and Zhang, Zhengyan and Liu, Zhiyuan and Han, Song and Sun, Maosong},
	year = {2024},
	note = {arXiv: 2402.04617},
}

@article{lu_controlled_2024,
	title = {A {Controlled} {Study} on {Long} {Context} {Extension} and {Generalization} in {LLMs}},
	volume = {abs/2409.12181},
	url = {https://doi.org/10.48550/arXiv.2409.12181},
	doi = {10.48550/ARXIV.2409.12181},
	journal = {CoRR},
	author = {Lu, Yi and Yan, Jing Nathan and Yang, Songlin and Chiu, Justin T. and Ren, Siyu and Yuan, Fei and Zhao, Wenting and Wu, Zhiyong and Rush, Alexander M.},
	year = {2024},
	note = {arXiv: 2409.12181},
}

@article{openai_gpt-4_2023,
	title = {{GPT}-4 {Technical} {Report}},
	volume = {abs/2303.08774},
	url = {https://doi.org/10.48550/arXiv.2303.08774},
	doi = {10.48550/ARXIV.2303.08774},
	journal = {CoRR},
	author = {{OpenAI}},
	year = {2023},
	note = {arXiv: 2303.08774},
}

@article{liu_lost_2024,
	title = {Lost in the {Middle}: {How} {Language} {Models} {Use} {Long} {Contexts}},
	volume = {12},
	url = {https://doi.org/10.1162/tacl\_a\_00638},
	doi = {10.1162/TACL_A_00638},
	journal = {Trans. Assoc. Comput. Linguistics},
	author = {Liu, Nelson F. and Lin, Kevin and Hewitt, John and Paranjape, Ashwin and Bevilacqua, Michele and Petroni, Fabio and Liang, Percy},
	year = {2024},
	pages = {157--173},
}

@inproceedings{vig_analyzing_2019,
	title = {Analyzing the {Structure} of {Attention} in a {Transformer} {Language} {Model}},
	url = {https://doi.org/10.18653/v1/W19-4808},
	doi = {10.18653/V1/W19-4808},
	booktitle = {Proceedings of the 2019 {ACL} {Workshop} {BlackboxNLP}: {Analyzing} and {Interpreting} {Neural} {Networks} for {NLP}, {BlackboxNLP}@{ACL} 2019, {Florence}, {Italy}, {August} 1, 2019},
	publisher = {Association for Computational Linguistics},
	author = {Vig, Jesse and Belinkov, Yonatan},
	editor = {Linzen, Tal and Chrupala, Grzegorz and Belinkov, Yonatan and Hupkes, Dieuwke},
	year = {2019},
	pages = {63--76},
}

@article{sun_transformer_2024,
	title = {Transformer {Layers} as {Painters}},
	volume = {abs/2407.09298},
	url = {https://doi.org/10.48550/arXiv.2407.09298},
	doi = {10.48550/ARXIV.2407.09298},
	journal = {CoRR},
	author = {Sun, Qi and Pickett, Marc and Nain, Aakash Kumar and Jones, Llion},
	year = {2024},
	note = {arXiv: 2407.09298},
}

@inproceedings{yang_hotpotqa_2018,
	title = {{HotpotQA}: {A} {Dataset} for {Diverse}, {Explainable} {Multi}-hop {Question} {Answering}},
	url = {https://doi.org/10.18653/v1/d18-1259},
	doi = {10.18653/V1/D18-1259},
	booktitle = {Proceedings of the 2018 {Conference} on {Empirical} {Methods} in {Natural} {Language} {Processing}, {Brussels}, {Belgium}, {October} 31 - {November} 4, 2018},
	publisher = {Association for Computational Linguistics},
	author = {Yang, Zhilin and Qi, Peng and Zhang, Saizheng and Bengio, Yoshua and Cohen, William W. and Salakhutdinov, Ruslan and Manning, Christopher D.},
	editor = {Riloff, Ellen and Chiang, David and Hockenmaier, Julia and Tsujii, Jun'ichi},
	year = {2018},
	pages = {2369--2380},
}

@inproceedings{bai_longbench_2024,
	title = {{LongBench}: {A} {Bilingual}, {Multitask} {Benchmark} for {Long} {Context} {Understanding}},
	url = {https://doi.org/10.18653/v1/2024.acl-long.172},
	doi = {10.18653/V1/2024.ACL-LONG.172},
	booktitle = {Proceedings of the 62nd {Annual} {Meeting} of the {Association} for {Computational} {Linguistics} ({Volume} 1: {Long} {Papers}), {ACL} 2024, {Bangkok}, {Thailand}, {August} 11-16, 2024},
	publisher = {Association for Computational Linguistics},
	author = {Bai, Yushi and Lv, Xin and Zhang, Jiajie and Lyu, Hongchang and Tang, Jiankai and Huang, Zhidian and Du, Zhengxiao and Liu, Xiao and Zeng, Aohan and Hou, Lei and Dong, Yuxiao and Tang, Jie and Li, Juanzi},
	editor = {Ku, Lun-Wei and Martins, Andre and Srikumar, Vivek},
	year = {2024},
	pages = {3119--3137},
}

@article{jiang_mistral_2023,
	title = {Mistral {7B}},
	volume = {abs/2310.06825},
	url = {https://doi.org/10.48550/arXiv.2310.06825},
	doi = {10.48550/ARXIV.2310.06825},
	journal = {CoRR},
	author = {Jiang, Albert Q. and Sablayrolles, Alexandre and Mensch, Arthur and Bamford, Chris and Chaplot, Devendra Singh and Casas, Diego de Las and Bressand, Florian and Lengyel, Gianna and Lample, Guillaume and Saulnier, Lucile and Lavaud, Lélio Renard and Lachaux, Marie-Anne and Stock, Pierre and Scao, Teven Le and Lavril, Thibaut and Wang, Thomas and Lacroix, Timothée and Sayed, William El},
	year = {2023},
	note = {arXiv: 2310.06825},
}

@article{dubey_llama_2024,
	title = {The {Llama} 3 {Herd} of {Models}},
	volume = {abs/2407.21783},
	url = {https://doi.org/10.48550/arXiv.2407.21783},
	doi = {10.48550/ARXIV.2407.21783},
	journal = {CoRR},
	author = {Dubey, Abhimanyu and Jauhri, Abhinav and Pandey, Abhinav and Kadian, Abhishek and Al-Dahle, Ahmad and Letman, Aiesha and Mathur, Akhil and Schelten, Alan and Yang, Amy and Fan, Angela and Goyal, Anirudh and Hartshorn, Anthony and Yang, Aobo and Mitra, Archi and Sravankumar, Archie and Korenev, Artem and Hinsvark, Arthur and Rao, Arun and Zhang, Aston and Rodriguez, Aurélien and Gregerson, Austen and Spataru, Ava and Rozière, Baptiste and Biron, Bethany and Tang, Binh and Chern, Bobbie and Caucheteux, Charlotte and Nayak, Chaya and Bi, Chloe and Marra, Chris and McConnell, Chris and Keller, Christian and Touret, Christophe and Wu, Chunyang and Wong, Corinne and Ferrer, Cristian Canton and Nikolaidis, Cyrus and Allonsius, Damien and Song, Daniel and Pintz, Danielle and Livshits, Danny and Esiobu, David and Choudhary, Dhruv and Mahajan, Dhruv and Garcia-Olano, Diego and Perino, Diego and Hupkes, Dieuwke and Lakomkin, Egor and AlBadawy, Ehab and Lobanova, Elina and Dinan, Emily and Smith, Eric Michael and Radenovic, Filip and Zhang, Frank and Synnaeve, Gabriel and Lee, Gabrielle and Anderson, Georgia Lewis and Nail, Graeme and Mialon, Grégoire and Pang, Guan and Cucurell, Guillem and Nguyen, Hailey and Korevaar, Hannah and Xu, Hu and Touvron, Hugo and Zarov, Iliyan and Ibarra, Imanol Arrieta and Kloumann, Isabel M. and Misra, Ishan and Evtimov, Ivan and Copet, Jade and Lee, Jaewon and Geffert, Jan and Vranes, Jana and Park, Jason and Mahadeokar, Jay and Shah, Jeet and Linde, Jelmer van der and Billock, Jennifer and Hong, Jenny and Lee, Jenya and Fu, Jeremy and Chi, Jianfeng and Huang, Jianyu and Liu, Jiawen and Wang, Jie and Yu, Jiecao and Bitton, Joanna and Spisak, Joe and Park, Jongsoo and Rocca, Joseph and Johnstun, Joshua and Saxe, Joshua and Jia, Junteng and Alwala, Kalyan Vasuden and Upasani, Kartikeya and Plawiak, Kate and Li, Ke and Heafield, Kenneth and Stone, Kevin and al, et},
	year = {2024},
	note = {arXiv: 2407.21783},
}

@inproceedings{zhang_mgte_2024,
	title = {{mGTE}: {Generalized} {Long}-{Context} {Text} {Representation} and {Reranking} {Models} for {Multilingual} {Text} {Retrieval}},
	url = {https://aclanthology.org/2024.emnlp-industry.103},
	booktitle = {Proceedings of the 2024 {Conference} on {Empirical} {Methods} in {Natural} {Language} {Processing}: {EMNLP} 2024 - {Industry} {Track}, {Miami}, {Florida}, {USA}, {November} 12-16, 2024},
	publisher = {Association for Computational Linguistics},
	author = {Zhang, Xin and Zhang, Yanzhao and Long, Dingkun and Xie, Wen and Dai, Ziqi and Tang, Jialong and Lin, Huan and Yang, Baosong and Xie, Pengjun and Huang, Fei and Zhang, Meishan and Li, Wenjie and Zhang, Min},
	editor = {Dernoncourt, Franck and Preotiuc-Pietro, Daniel and Shimorina, Anastasia},
	year = {2024},
	pages = {1393--1412},
}

@inproceedings{lewis_retrieval-augmented_2020,
	title = {Retrieval-{Augmented} {Generation} for {Knowledge}-{Intensive} {NLP} {Tasks}},
	url = {https://proceedings.neurips.cc/paper/2020/hash/6b493230205f780e1bc26945df7481e5-Abstract.html},
	booktitle = {Advances in {Neural} {Information} {Processing} {Systems} 33: {Annual} {Conference} on {Neural} {Information} {Processing} {Systems} 2020, {NeurIPS} 2020, {December} 6-12, 2020, virtual},
	author = {Lewis, Patrick S. H. and Perez, Ethan and Piktus, Aleksandra and Petroni, Fabio and Karpukhin, Vladimir and Goyal, Naman and Küttler, Heinrich and Lewis, Mike and Yih, Wen-tau and Rocktäschel, Tim and Riedel, Sebastian and Kiela, Douwe},
	editor = {Larochelle, Hugo and Ranzato, Marc'Aurelio and Hadsell, Raia and Balcan, Maria-Florina and Lin, Hsuan-Tien},
	year = {2020},
}

@inproceedings{karpukhin_dense_2020,
	title = {Dense {Passage} {Retrieval} for {Open}-{Domain} {Question} {Answering}},
	url = {https://doi.org/10.18653/v1/2020.emnlp-main.550},
	doi = {10.18653/V1/2020.EMNLP-MAIN.550},
	booktitle = {Proceedings of the 2020 {Conference} on {Empirical} {Methods} in {Natural} {Language} {Processing}, {EMNLP} 2020, {Online}, {November} 16-20, 2020},
	publisher = {Association for Computational Linguistics},
	author = {Karpukhin, Vladimir and Oguz, Barlas and Min, Sewon and Lewis, Patrick S. H. and Wu, Ledell and Edunov, Sergey and Chen, Danqi and Yih, Wen-tau},
	editor = {Webber, Bonnie and Cohn, Trevor and He, Yulan and Liu, Yang},
	year = {2020},
	pages = {6769--6781},
}

@article{morris_contextual_2024,
	title = {Contextual {Document} {Embeddings}},
	volume = {abs/2410.02525},
	url = {https://doi.org/10.48550/arXiv.2410.02525},
	doi = {10.48550/ARXIV.2410.02525},
	journal = {CoRR},
	author = {Morris, John X. and Rush, Alexander M.},
	year = {2024},
	note = {arXiv: 2410.02525},
}

@inproceedings{xiong_approximate_2021,
	title = {Approximate {Nearest} {Neighbor} {Negative} {Contrastive} {Learning} for {Dense} {Text} {Retrieval}},
	url = {https://openreview.net/forum?id=zeFrfgyZln},
	booktitle = {9th {International} {Conference} on {Learning} {Representations}, {ICLR} 2021, {Virtual} {Event}, {Austria}, {May} 3-7, 2021},
	publisher = {OpenReview.net},
	author = {Xiong, Lee and Xiong, Chenyan and Li, Ye and Tang, Kwok-Fung and Liu, Jialin and Bennett, Paul N. and Ahmed, Junaid and Overwijk, Arnold},
	year = {2021},
}

@inproceedings{ni_large_2022,
	title = {Large {Dual} {Encoders} {Are} {Generalizable} {Retrievers}},
	url = {https://doi.org/10.18653/v1/2022.emnlp-main.669},
	doi = {10.18653/V1/2022.EMNLP-MAIN.669},
	booktitle = {Proceedings of the 2022 {Conference} on {Empirical} {Methods} in {Natural} {Language} {Processing}, {EMNLP} 2022, {Abu} {Dhabi}, {United} {Arab} {Emirates}, {December} 7-11, 2022},
	publisher = {Association for Computational Linguistics},
	author = {Ni, Jianmo and Qu, Chen and Lu, Jing and Dai, Zhuyun and Ábrego, Gustavo Hernández and Ma, Ji and Zhao, Vincent Y. and Luan, Yi and Hall, Keith B. and Chang, Ming-Wei and Yang, Yinfei},
	editor = {Goldberg, Yoav and Kozareva, Zornitsa and Zhang, Yue},
	year = {2022},
	pages = {9844--9855},
}

@inproceedings{bahdanau_neural_2015,
	title = {Neural {Machine} {Translation} by {Jointly} {Learning} to {Align} and {Translate}},
	url = {http://arxiv.org/abs/1409.0473},
	booktitle = {3rd {International} {Conference} on {Learning} {Representations}, {ICLR} 2015, {San} {Diego}, {CA}, {USA}, {May} 7-9, 2015, {Conference} {Track} {Proceedings}},
	author = {Bahdanau, Dzmitry and Cho, Kyunghyun and Bengio, Yoshua},
	editor = {Bengio, Yoshua and LeCun, Yann},
	year = {2015},
}

@article{shang_longrope2_2025,
	title = {{LongRoPE2}: {Near}-{Lossless} {LLM} {Context} {Window} {Scaling}},
	volume = {abs/2502.20082},
	url = {https://doi.org/10.48550/arXiv.2502.20082},
	doi = {10.48550/ARXIV.2502.20082},
	journal = {CoRR},
	author = {Shang, Ning and Zhang, Li Lyna and Wang, Siyuan and Zhang, Gaokai and Lopez, Gilsinia and Yang, Fan and Chen, Weizhu and Yang, Mao},
	year = {2025},
	note = {arXiv: 2502.20082},
}

@inproceedings{ho_constructing_2020,
	title = {Constructing {A} {Multi}-hop {QA} {Dataset} for {Comprehensive} {Evaluation} of {Reasoning} {Steps}},
	url = {https://doi.org/10.18653/v1/2020.coling-main.580},
	doi = {10.18653/V1/2020.COLING-MAIN.580},
	booktitle = {Proceedings of the 28th {International} {Conference} on {Computational} {Linguistics}, {COLING} 2020, {Barcelona}, {Spain} ({Online}), {December} 8-13, 2020},
	publisher = {International Committee on Computational Linguistics},
	author = {Ho, Xanh and Nguyen, Anh-Khoa Duong and Sugawara, Saku and Aizawa, Akiko},
	editor = {Scott, Donia and Bel, Núria and Zong, Chengqing},
	year = {2020},
	pages = {6609--6625},
}

@article{trivedi_musique_2022,
	title = {{\textbackslash}unicode9835 {MuSiQue}: {Multihop} {Questions} via {Single}-hop {Question} {Composition}},
	volume = {10},
	url = {https://doi.org/10.1162/tacl\_a\_00475},
	doi = {10.1162/TACL_A_00475},
	journal = {Trans. Assoc. Comput. Linguistics},
	author = {Trivedi, Harsh and Balasubramanian, Niranjan and Khot, Tushar and Sabharwal, Ashish},
	year = {2022},
	pages = {539--554},
}

@techreport{openai_gpt-5_2025,
	title = {{GPT}-5 {System} {Card}},
	url = {https://cdn.openai.com/gpt-5-system-card.pdf},
	author = {{OpenAI}},
	year = {2025},
}

@inproceedings{dong_mc-indexing_2024,
	title = {{MC}-indexing: {Effective} {Long} {Document} {Retrieval} via {Multi}-view {Content}-aware {Indexing}},
	url = {https://doi.org/10.18653/v1/2024.findings-emnlp.150},
	doi = {10.18653/V1/2024.FINDINGS-EMNLP.150},
	booktitle = {Findings of the {Association} for {Computational} {Linguistics}: {EMNLP} 2024, {Miami}, {Florida}, {USA}, {November} 12-16, 2024},
	publisher = {Association for Computational Linguistics},
	author = {Dong, Kuicai and Deik, Derrick-Goh-Xin and Lee, Yi and Zhang, Hao and Li, Xiangyang and Zhang, Cong and Liu, Yong},
	editor = {Al-Onaizan, Yaser and Bansal, Mohit and Chen, Yun-Nung},
	year = {2024},
	pages = {2673--2691},
}

@inproceedings{muennighoff_generative_2025,
	title = {Generative {Representational} {Instruction} {Tuning}},
	url = {https://openreview.net/forum?id=BC4lIvfSzv},
	booktitle = {The {Thirteenth} {International} {Conference} on {Learning} {Representations}, {ICLR} 2025, {Singapore}, {April} 24-28, 2025},
	publisher = {OpenReview.net},
	author = {Muennighoff, Niklas and Su, Hongjin and Wang, Liang and Yang, Nan and Wei, Furu and Yu, Tao and Singh, Amanpreet and Kiela, Douwe},
	year = {2025},
}

@article{gunther_late_2024,
	title = {Late {Chunking}: {Contextual} {Chunk} {Embeddings} {Using} {Long}-{Context} {Embedding} {Models}},
	volume = {abs/2409.04701},
	url = {https://doi.org/10.48550/arXiv.2409.04701},
	doi = {10.48550/ARXIV.2409.04701},
	journal = {CoRR},
	author = {Günther, Michael and Mohr, Isabelle and Wang, Bo and Xiao, Han},
	year = {2024},
	note = {arXiv: 2409.04701},
}

@article{conti_context_2025,
	title = {Context is {Gold} to find the {Gold} {Passage}: {Evaluating} and {Training} {Contextual} {Document} {Embeddings}},
	volume = {abs/2505.24782},
	url = {https://doi.org/10.48550/arXiv.2505.24782},
	doi = {10.48550/ARXIV.2505.24782},
	journal = {CoRR},
	author = {Conti, Max and Faysse, Manuel and Viaud, Gautier and Bosselut, Antoine and Hudelot, Céline and Colombo, Pierre},
	year = {2025},
	note = {arXiv: 2505.24782},
}

@inproceedings{kwon_efficient_2023,
	title = {Efficient {Memory} {Management} for {Large} {Language} {Model} {Serving} with {PagedAttention}},
	booktitle = {Proceedings of the {ACM} {SIGOPS} 29th {Symposium} on {Operating} {Systems} {Principles}},
	author = {Kwon, Woosuk and Li, Zhuohan and Zhuang, Siyuan and Sheng, Ying and Zheng, Lianmin and Yu, Cody Hao and Gonzalez, Joseph E. and Zhang, Hao and Stoica, Ion},
	year = {2023},
}

@article{yang_qwen25_2024,
	title = {Qwen2.5 {Technical} {Report}},
	volume = {abs/2412.15115},
	url = {https://doi.org/10.48550/arXiv.2412.15115},
	doi = {10.48550/ARXIV.2412.15115},
	journal = {CoRR},
	author = {Yang, An and Yang, Baosong and Zhang, Beichen and Hui, Binyuan and Zheng, Bo and Yu, Bowen and Li, Chengyuan and Liu, Dayiheng and Huang, Fei and Wei, Haoran and Lin, Huan and Yang, Jian and Tu, Jianhong and Zhang, Jianwei and Yang, Jianxin and Yang, Jiaxi and Zhou, Jingren and Lin, Junyang and Dang, Kai and Lu, Keming and Bao, Keqin and Yang, Kexin and Yu, Le and Li, Mei and Xue, Mingfeng and Zhang, Pei and Zhu, Qin and Men, Rui and Lin, Runji and Li, Tianhao and Xia, Tingyu and Ren, Xingzhang and Ren, Xuancheng and Fan, Yang and Su, Yang and Zhang, Yichang and Wan, Yu and Liu, Yuqiong and Cui, Zeyu and Zhang, Zhenru and Qiu, Zihan},
	year = {2024},
	note = {arXiv: 2412.15115},
}

@inproceedings{willette_training_2025,
	title = {Training {Free} {Exponential} {Context} {Extension} via {Cascading} {KV} {Cache}},
	url = {https://openreview.net/forum?id=dSneEp59yX},
	booktitle = {The {Thirteenth} {International} {Conference} on {Learning} {Representations}, {ICLR} 2025, {Singapore}, {April} 24-28, 2025},
	publisher = {OpenReview.net},
	author = {Willette, Jeffrey and Lee, Heejun and Lee, Youngwan and Jeon, Myeongjae and Hwang, Sung Ju},
	year = {2025},
}

@inproceedings{an_training-free_2024,
	title = {Training-{Free} {Long}-{Context} {Scaling} of {Large} {Language} {Models}},
	url = {https://openreview.net/forum?id=If4xW9vF7U},
	booktitle = {Forty-first {International} {Conference} on {Machine} {Learning}, {ICML} 2024, {Vienna}, {Austria}, {July} 21-27, 2024},
	publisher = {OpenReview.net},
	author = {An, Chenxin and Huang, Fei and Zhang, Jun and Gong, Shansan and Qiu, Xipeng and Zhou, Chang and Kong, Lingpeng},
	year = {2024},
}

@inproceedings{liu_reattention_2025,
	title = {{ReAttention}: {Training}-{Free} {Infinite} {Context} with {Finite} {Attention} {Scope}},
	url = {https://openreview.net/forum?id=KDGP8yAz5b},
	booktitle = {The {Thirteenth} {International} {Conference} on {Learning} {Representations}, {ICLR} 2025, {Singapore}, {April} 24-28, 2025},
	publisher = {OpenReview.net},
	author = {Liu, Xiaoran and Li, Ruixiao and Liu, Zhigeng and Guo, Qipeng and Song, Yuerong and Lv, Kai and Yan, Hang and Li, Linlin and Liu, Qun and Qiu, Xipeng},
	year = {2025},
}

@inproceedings{jin_llm_2024,
	title = {{LLM} {Maybe} {LongLM}: {SelfExtend} {LLM} {Context} {Window} {Without} {Tuning}},
	url = {https://openreview.net/forum?id=nkOMLBIiI7},
	booktitle = {Forty-first {International} {Conference} on {Machine} {Learning}, {ICML} 2024, {Vienna}, {Austria}, {July} 21-27, 2024},
	publisher = {OpenReview.net},
	author = {Jin, Hongye and Han, Xiaotian and Yang, Jingfeng and Jiang, Zhimeng and Liu, Zirui and Chang, Chia-Yuan and Chen, Huiyuan and Hu, Xia},
	year = {2024},
}

@inproceedings{xu_extending_2025,
	title = {Extending {LLM} {Context} {Window} with {Adaptive} {Grouped} {Positional} {Encoding}: {A} {Training}-{Free} {Method}},
	url = {https://aclanthology.org/2025.acl-long.28/},
	booktitle = {Proceedings of the 63rd {Annual} {Meeting} of the {Association} for {Computational} {Linguistics} ({Volume} 1: {Long} {Papers}), {ACL} 2025, {Vienna}, {Austria}, {July} 27 - {August} 1, 2025},
	publisher = {Association for Computational Linguistics},
	author = {Xu, Xinhao and Li, Jiaxin and Chen, Hui and Lin, Zijia and Han, Jungong and Ding, Guiguang},
	editor = {Che, Wanxiang and Nabende, Joyce and Shutova, Ekaterina and Pilehvar, Mohammad Taher},
	year = {2025},
	pages = {573--587},
}

@inproceedings{wang_dapr_2024,
	title = {{DAPR}: {A} {Benchmark} on {Document}-{Aware} {Passage} {Retrieval}},
	url = {https://doi.org/10.18653/v1/2024.acl-long.236},
	doi = {10.18653/V1/2024.ACL-LONG.236},
	booktitle = {Proceedings of the 62nd {Annual} {Meeting} of the {Association} for {Computational} {Linguistics} ({Volume} 1: {Long} {Papers}), {ACL} 2024, {Bangkok}, {Thailand}, {August} 11-16, 2024},
	publisher = {Association for Computational Linguistics},
	author = {Wang, Kexin and Reimers, Nils and Gurevych, Iryna},
	editor = {Ku, Lun-Wei and Martins, Andre and Srikumar, Vivek},
	year = {2024},
	pages = {4313--4330},
}

@article{ye_infinite_2025,
	title = {Infinite {Retrieval}: {Attention} {Enhanced} {LLMs} in {Long}-{Context} {Processing}},
	volume = {abs/2502.12962},
	url = {https://doi.org/10.48550/arXiv.2502.12962},
	doi = {10.48550/ARXIV.2502.12962},
	journal = {CoRR},
	author = {Ye, Xiaoju and Wang, Zhichun and Wang, Jingyuan},
	year = {2025},
	note = {arXiv: 2502.12962},
}

@article{zhang_qwen3_2025,
	title = {Qwen3 {Embedding}: {Advancing} {Text} {Embedding} and {Reranking} {Through} {Foundation} {Models}},
	volume = {abs/2506.05176},
	url = {https://doi.org/10.48550/arXiv.2506.05176},
	doi = {10.48550/ARXIV.2506.05176},
	journal = {CoRR},
	author = {Zhang, Yanzhao and Li, Mingxin and Long, Dingkun and Zhang, Xin and Lin, Huan and Yang, Baosong and Xie, Pengjun and Yang, An and Liu, Dayiheng and Lin, Junyang and Huang, Fei and Zhou, Jingren},
	year = {2025},
	note = {arXiv: 2506.05176},
}

@inproceedings{lee_qasa_2023,
	series = {Proceedings of {Machine} {Learning} {Research}},
	title = {{QASA}: {Advanced} {Question} {Answering} on {Scientific} {Articles}},
	volume = {202},
	url = {https://proceedings.mlr.press/v202/lee23n.html},
	booktitle = {International {Conference} on {Machine} {Learning}, {ICML} 2023, 23-29 {July} 2023, {Honolulu}, {Hawaii}, {USA}},
	publisher = {PMLR},
	author = {Lee, Yoonjoo and Lee, Kyungjae and Park, Sunghyun and Hwang, Dasol and Kim, Jaehyeon and Lee, Hong-In and Lee, Moontae},
	editor = {Krause, Andreas and Brunskill, Emma and Cho, Kyunghyun and Engelhardt, Barbara and Sabato, Sivan and Scarlett, Jonathan},
	year = {2023},
	pages = {19036--19052},
}

@inproceedings{dasigi_dataset_2021,
	title = {A {Dataset} of {Information}-{Seeking} {Questions} and {Answers} {Anchored} in {Research} {Papers}},
	url = {https://doi.org/10.18653/v1/2021.naacl-main.365},
	doi = {10.18653/V1/2021.NAACL-MAIN.365},
	booktitle = {Proceedings of the 2021 {Conference} of the {North} {American} {Chapter} of the {Association} for {Computational} {Linguistics}: {Human} {Language} {Technologies}, {NAACL}-{HLT} 2021, {Online}, {June} 6-11, 2021},
	publisher = {Association for Computational Linguistics},
	author = {Dasigi, Pradeep and Lo, Kyle and Beltagy, Iz and Cohan, Arman and Smith, Noah A. and Gardner, Matt},
	editor = {Toutanova, Kristina and Rumshisky, Anna and Zettlemoyer, Luke and Hakkani-Tür, Dilek and Beltagy, Iz and Bethard, Steven and Cotterell, Ryan and Chakraborty, Tanmoy and Zhou, Yichao},
	year = {2021},
	pages = {4599--4610},
}

@inproceedings{monteiro_repliqa_2024,
	title = {{RepLiQA}: {A} {Question}-{Answering} {Dataset} for {Benchmarking} {LLMs} on {Unseen} {Reference} {Content}},
	url = {http://papers.nips.cc/paper\_files/paper/2024/hash/2b23626015b6311369e95a70735cbb72-Abstract-Datasets\_and\_Benchmarks\_Track.html},
	booktitle = {Advances in {Neural} {Information} {Processing} {Systems} 38: {Annual} {Conference} on {Neural} {Information} {Processing} {Systems} 2024, {NeurIPS} 2024, {Vancouver}, {BC}, {Canada}, {December} 10 - 15, 2024},
	author = {Monteiro, João and Noël, Pierre-André and Marcotte, Étienne and Mudumba, Sai Rajeswar and Zantedeschi, Valentina and Vázquez, David and Chapados, Nicolas and Pal, Chris and Taslakian, Perouz},
	editor = {Globersons, Amir and Mackey, Lester and Belgrave, Danielle and Fan, Angela and Paquet, Ulrich and Tomczak, Jakub M. and Zhang, Cheng},
	year = {2024},
}

@inproceedings{bai_longbench_2025,
	title = {{LongBench} v2: {Towards} {Deeper} {Understanding} and {Reasoning} on {Realistic} {Long}-context {Multitasks}},
	url = {https://aclanthology.org/2025.acl-long.183/},
	booktitle = {Proceedings of the 63rd {Annual} {Meeting} of the {Association} for {Computational} {Linguistics} ({Volume} 1: {Long} {Papers}), {ACL} 2025, {Vienna}, {Austria}, {July} 27 - {August} 1, 2025},
	publisher = {Association for Computational Linguistics},
	author = {Bai, Yushi and Tu, Shangqing and Zhang, Jiajie and Peng, Hao and Wang, Xiaozhi and Lv, Xin and Cao, Shulin and Xu, Jiazheng and Hou, Lei and Dong, Yuxiao and Tang, Jie and Li, Juanzi},
	editor = {Che, Wanxiang and Nabende, Joyce and Shutova, Ekaterina and Pilehvar, Mohammad Taher},
	year = {2025},
	pages = {3639--3664},
}

@article{cao_single-pass_2025,
	title = {Single-{Pass} {Document} {Scanning} for {Question} {Answering}},
	volume = {abs/2504.03101},
	url = {https://doi.org/10.48550/arXiv.2504.03101},
	doi = {10.48550/ARXIV.2504.03101},
	journal = {CoRR},
	author = {Cao, Weili and Wang, Jianyou and Zheng, Youze and Bao, Longtian and Zheng, Qirui and Berg-Kirkpatrick, Taylor and Paturi, Ramamohan and Bergen, Leon},
	year = {2025},
	note = {arXiv: 2504.03101},
}

@misc{noauthor_gkamradtllmtest_needleinahaystack_nodate,
    author = {gkamradt},
	title = {gkamradt/{LLMTest}\_NeedleInAHaystack: {Doing} simple retrieval from {LLM} models at various context lengths to measure accuracy},
	url = {https://github.com/gkamradt/LLMTest_NeedleInAHaystack},
	urldate = {2026-01-06},
}

\appendix

\section{Details of LongBench-v2-Retrieval}
\label{appendix:dataset}

\subsection{Comparison with Existing Long Document Retrieval Datasets}
\begin{table*}[!htb]
    \tiny
    \resizebox{\textwidth}{!}{
        \begin{tabular}{lcccccc}
            \toprule
            \textbf{Dataset} & \textbf{Size} & \textbf{Avg Length} & \textbf{Max Length} & \textbf{Avg Paragraph Count} & \textbf{Max Paragraph Count} & \textbf{Avg Number of Evidences} \\
            \midrule
            QASA \cite{lee_qasa_2023} & 518 & 4665.09 & 15796 & 52.45 & 186 & 1.55 \\
            Qasper \cite{dasigi_dataset_2021} & 1307 & 3442.26 & 21043 & 61.94 & 328 & 1.75 \\
            RepLiQA \cite{monteiro_repliqa_2024} & 89770 & 970.50 & 1799 & 22.31 & 58 & 1.00 \\
            DAPR (ConditionalQA) \cite{wang_dapr_2024} & 2517 & 1298.13 & 7733 & 105.04 & 559 & 4.08 \\
            DAPR (MS MARCO) & 98466 & 1059.75 & 152515 & 4.05 & 43 & 1.03 \\
            DAPR (NaturalQuestions) & 7220 & 2438.86 & 28871 & 30.14 & 535 & 1.11 \\
            LongBench-v2-Retrieval (Ours) & 140 & 106025.49 & 350135 & 2460.23 & 13944 & 2.59 \\
            \bottomrule
        \end{tabular}
    }
    \caption{Comparison of long document retrieval retrieval datasets. "Average Length" and "Maximum Length" are measured in number of words. For datasets consisting of multiple splits, we count the total number of samples across all splits.}
    \label{table:dataset-stat}
\end{table*}

Table \ref{table:dataset-stat} compares the lengths and number of evidences of existing long document retrieval datasets with our newly constructed dataset. The average document length of our dataset is significantly larger than other existing datasets, while the average number of evidences is also larger than most datasets we compared with.

\subsection{Example Queries}

The following are example queries for each type in our constructed dataset:

\begin{tcolorbox}[title={Example Single-hop Query}]
Question: According to the Report of the Independent High-Level Expert Group on Climate Finance, what is the estimated annual spend on the transformation of the energy system and investing in sustainable agriculture for emerging markets and developing countries by 2030?

Answer: Around \$2.4 trillion per year.
\end{tcolorbox}

\begin{tcolorbox}[title={Example Comparison Query}]
Question: Did Czechia receive more funding under EU cohesion policy in 2014-2020 compared to 2007-2013?

Answer: No
\end{tcolorbox}

\begin{tcolorbox}[title={Example Composition Query}]
Question: What are the outstanding health features of the product that had over 2.5 million shipments within three months of its launch?

Answer: Improved Activity Rings and the Stay Fit app to help users build healthier lifestyles.
\end{tcolorbox}

\begin{tcolorbox}[title={Example Summarization Query}]
Question: What are the current challenges of dynamic virtual cluster provisioning in geo-distributed clouds as outlined in the paper?

Answer: There is not yet an efficient resource allocation algorithm for VC provisioning even in the offline case with all user VC requests known, and an efficient pricing mechanism to charge users for the VCs on the go is missing. Moreover, online auction of an entire virtual cluster, including VMs and the network in-between, has not been studied.
\end{tcolorbox}

\section{Detailed Attention Analysis}
\label{appendix:analysis}

\begin{figure}[!h]
    \centering
    \includegraphics[width=\linewidth]{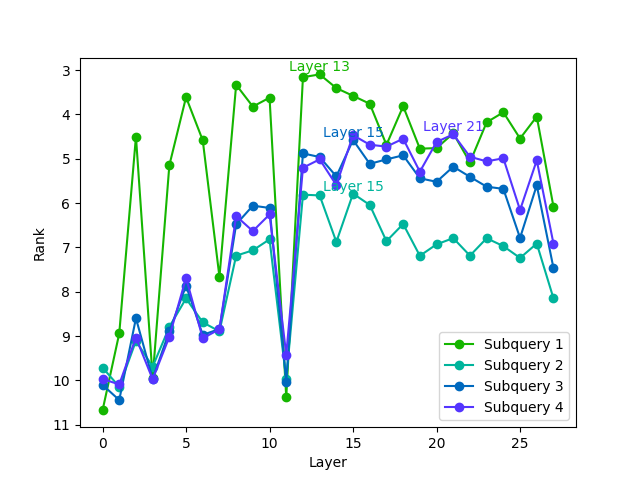}
    \caption{Average ranks of the gold paragraph for each subquery over all queries in the dataset for LLaMA-3.2 3B. The layer achieving the highest average rank for each subquery is marked at the top of each line. }    
    \label{fig:a-llama}
\end{figure}

\begin{figure}[!h]
    \centering
    \includegraphics[width=\linewidth]{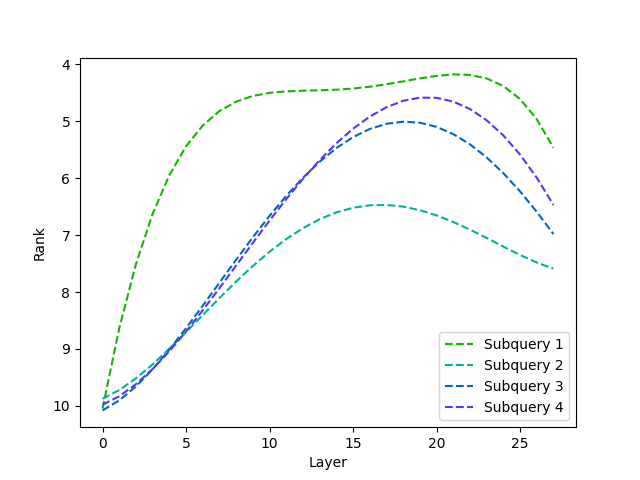}
    \caption{The quartic approximation of the average ranks in Figure \ref{fig:a-llama}.}
    \label{fig:b-llama}
\end{figure}

\begin{figure}[!h]
    \centering
    \includegraphics[width=\linewidth]{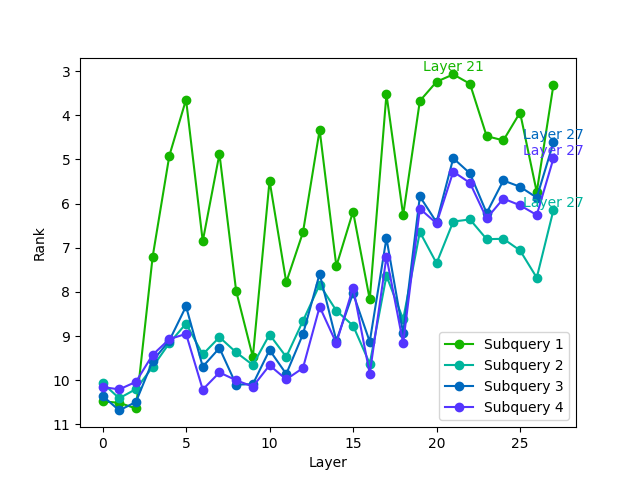}
    \caption{Average ranks of the gold paragraph for each subquery over all queries in the dataset for Qwen-2.5 3B. The layer achieving the highest average rank for each subquery is marked at the top of each line. }   
    \label{fig:a-qwen}
\end{figure}

\begin{figure}[!h]
    \centering
    \includegraphics[width=\linewidth]{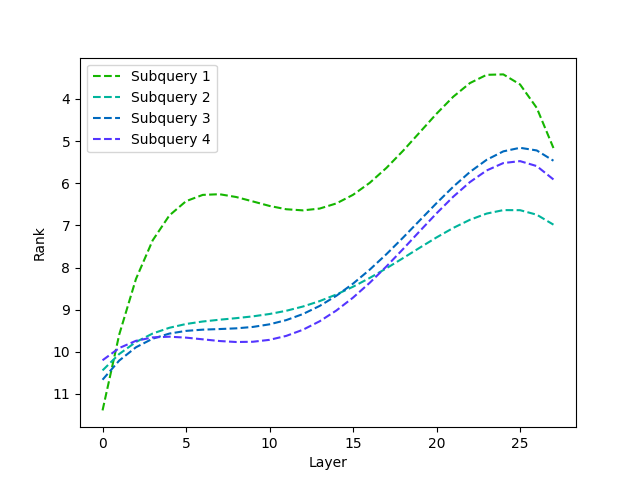}
    \caption{The quartic approximation of the average ranks in Figure \ref{fig:a-qwen}.}
    \label{fig:b-qwen}
\end{figure}

\begin{figure}[!h]
    \centering
    \includegraphics[width=\linewidth]{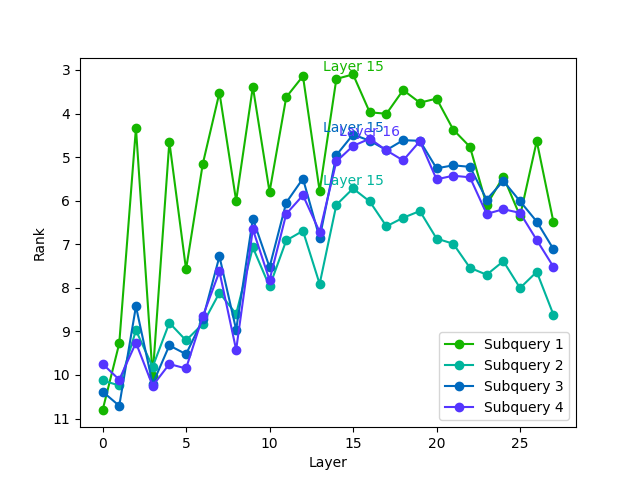}
    \caption{Average ranks of the gold paragraph for each subquery over all queries in the dataset for Mistral 7B. The layer achieving the highest average rank for each subquery is marked at the top of each line. }    
    \label{fig:a-mistral}
\end{figure}

\begin{figure}[!h]
    \centering
    \includegraphics[width=\linewidth]{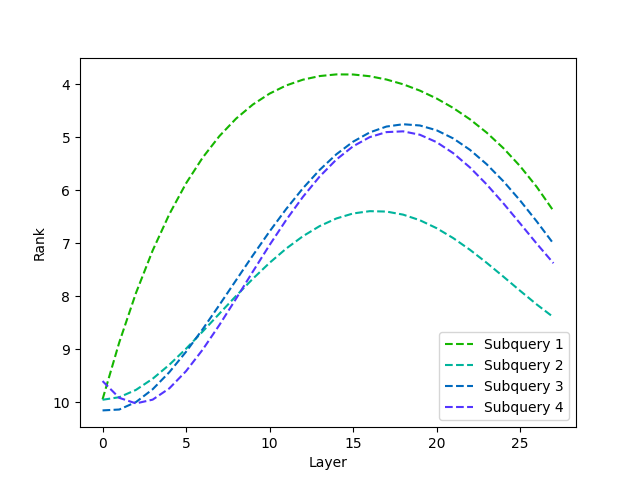}
    \caption{The quartic approximation of the average ranks in Figure \ref{fig:a-mistral}.}
    \label{fig:b-mistral}
\end{figure}

Figures \ref{fig:a-llama}, \ref{fig:a-qwen}, and \ref{fig:a-mistral} show the average ranks of gold paragraphs over all queries, while Figures \ref{fig:b-llama}, \ref{fig:b-qwen}, and \ref{fig:b-mistral} plot the corresponding quartic approximations for LLaMA, Qwen, and Mistral models, respectively. We found our conclusions in Section \ref{sec:observation} generally holds for all three LLMs we analyzed. 

\begin{figure}[!t]
    \centering
    \includegraphics[width=\linewidth]{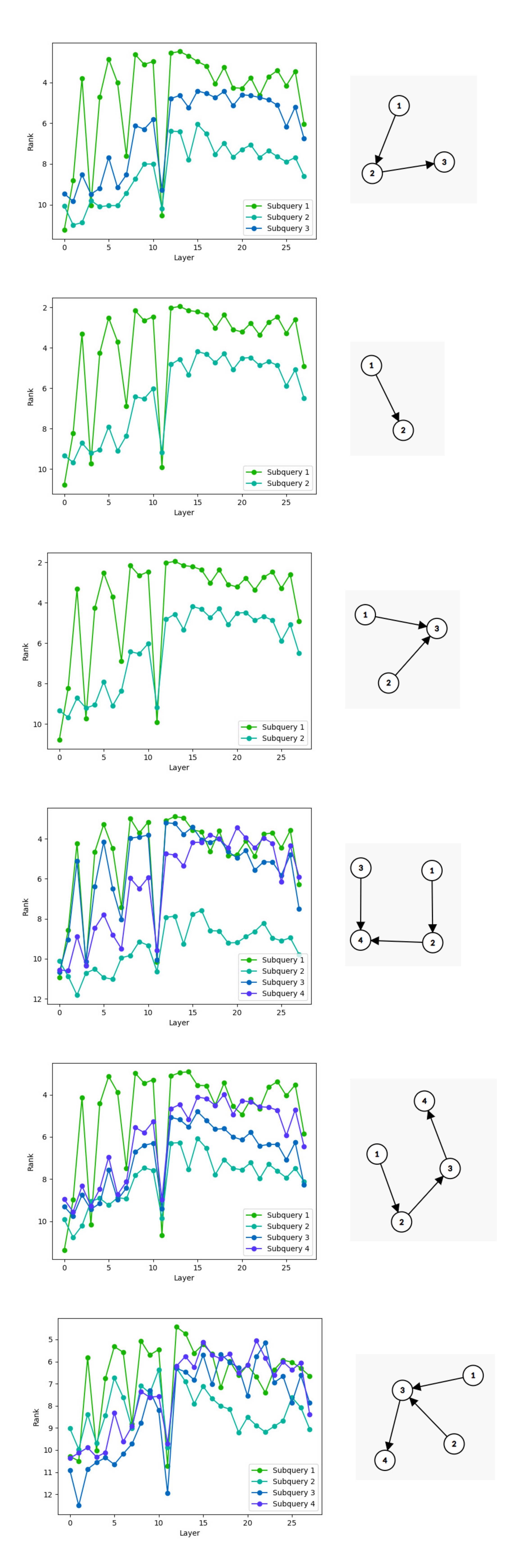}
    \caption{Detailed average ranks for different types of queries.}
    \label{fig:c-llama}
\end{figure}

\begin{figure}[!t]
    \centering
    \includegraphics[width=\linewidth]{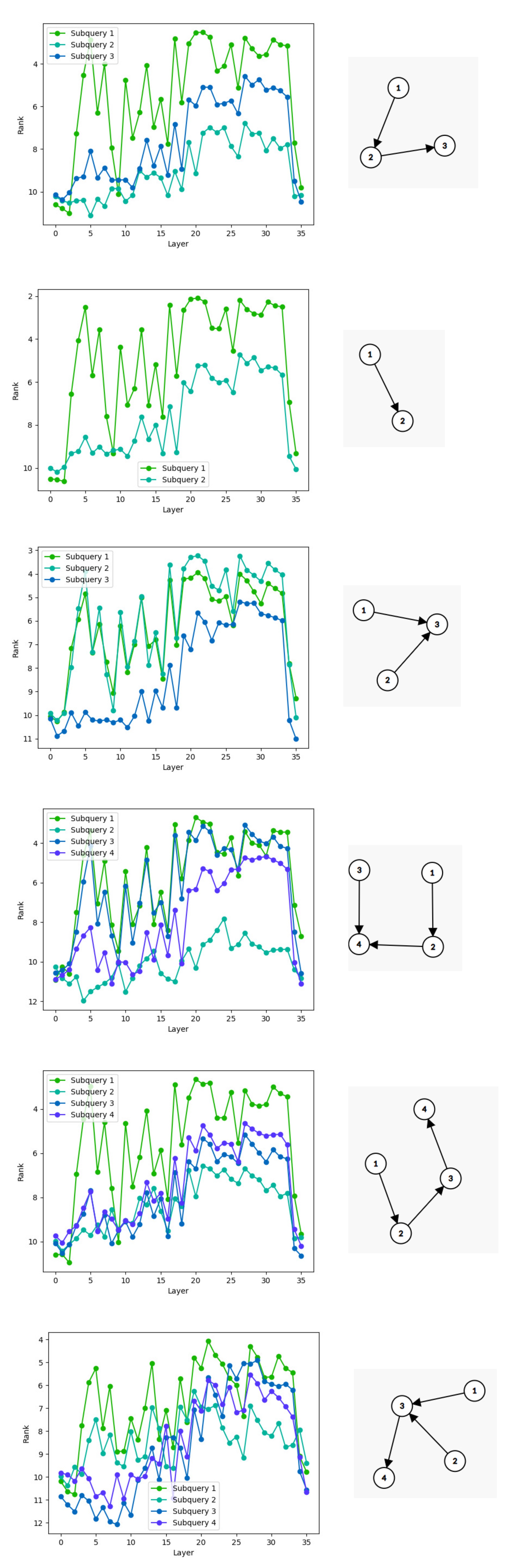}
    \caption{Detailed average ranks for different types of queries.}
    \label{fig:c-qwen}
\end{figure}

\begin{figure}[!t]
    \centering
    \includegraphics[width=\linewidth]{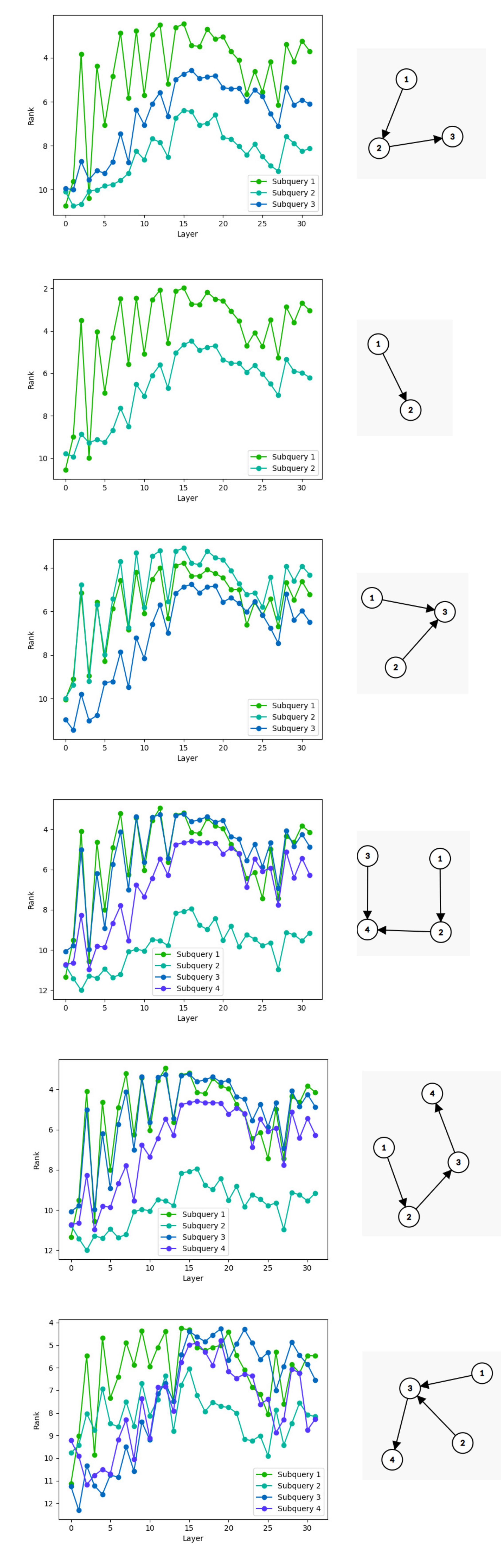}
    \caption{Detailed average ranks for different types of queries.}
    \label{fig:c-mistral}
\end{figure}

Figures \ref{fig:c-llama}, \ref{fig:c-qwen}, and \ref{fig:c-mistral} also present the detailed results for each type of queries in the dataset. The graph attached to the right of each figure explains the dependencies of the type, where $a\rightarrow b$ denotes subquery $b$ depends on the answer to subquery $a$. These results further validated our claim that LLMs adjust the token embeddings through the layers with additional contextual information, since LLMs tend to focus on subqueries depending on other subqueries in later layers so that necessary information has been collected before the layers. 





\newpage

\section{Full Evaluation Results on Retrieval Datasets}
\label{appendix:retrieval}

The full evaluation results on retrieval datasets with different values of \texttt{top\_k} can be found in Table \ref{table:retrieval-result-single-full}. Our proposed method is able to consistently outperform the baseline methods across all values of \texttt{top\_k} we selected. 

\begin{table*}[!htb]
    \centering
    \tiny
    \resizebox{\textwidth}{!}{
        \begin{tabular}{l|ccccccc|c}
            \toprule
            & QASA & Qasper & RepLiQA & ConditionalQA & NaturalQuestions & LongBench-v2-Retrieval & Average \\
            Average Length & 4665.09 & 3442.26 & 970.50 & 1298.13 & 2548.97 & 106025.49 & \\
            \midrule
            \textbf{Sparse Models} & & & & & & & \\
            \quad BM25 & & & & & & & \\
            \quad \quad \texttt{top\_k=1} & 0.4530 & 0.1983 & 0.7862 & 0.1412 & 0.3174 & 0.3576 & 0.3756 \\
            \quad \quad \texttt{top\_k=2} & 0.4255 & 0.2260 & 0.6133 & 0.1714 & 0.3284 & 0.3296 & 0.3490 \\
            \quad \quad \texttt{top\_k=3} & 0.3795 & 0.2304 & 0.4896 & 0.1988 & 0.3055 & 0.3126 & 0.3194 \\
            \quad \quad \texttt{top\_k=5} & 0.3087 & 0.2101 & 0.3475 & 0.2044 & 0.2572 & 0.2703 & 0.2665 \\
            \midrule
            \textbf{Dense Models} & & & & & & & \\
            \quad DPR & & & & & & & \\
            \quad \quad \texttt{top\_k=1} & 0.1979 & 0.0326 & 0.4655 & 0.1189 & 0.2743 & 0.0900 & 0.1965 \\
            \quad \quad \texttt{top\_k=2} & 0.2138 & 0.0462 & 0.4177 & 0.1470 & 0.3028 & 0.1139 & 0.2069 \\
            \quad \quad \texttt{top\_k=3} & 0.2181 & 0.0529 & 0.3602 & 0.1542 & 0.2981 & 0.1226 & 0.2010 \\
            \quad \quad \texttt{top\_k=5} & 0.2055 & 0.0608 & 0.2827 & 0.1689 & 0.2694 & 0.1073 & 0.1824 \\
            \quad ANCE & & & & & & & \\
            \quad \quad \texttt{top\_k=1} & 0.4216 & 0.2886 & 0.7364 & 0.1804 & 0.4309 & 0.3269 & 0.3975 \\
            \quad \quad \texttt{top\_k=2} & 0.4028 & 0.2957 & 0.5826 & 0.1898 & 0.4223 & 0.3500 & 0.3739 \\
            \quad \quad \texttt{top\_k=3} & 0.3724 & 0.2856 & 0.4686 & 0.1893 & 0.3818 & 0.3240 & 0.3370 \\
            \quad \quad \texttt{top\_k=5} & 0.3019 & 0.2515 & 0.3372 & 0.1945 & 0.3084 & 0.2584 & 0.2753 \\
            \quad CDE & & & & & & & \\
            \quad \quad \texttt{top\_k=1} & 0.1061 & 0.0810 & 0.2535 & 0.1025 & 0.2117 & 0.0452 & 0.1334 \\
            \quad \quad \texttt{top\_k=2} & 0.1199 & 0.0969 & 0.2598 & 0.1419 & 0.2398 & 0.0443 & 0.1504 \\
            \quad \quad \texttt{top\_k=3} & 0.1341 & 0.0962 & 0.2449 & 0.1547 & 0.2416 & 0.0485 & 0.1532 \\
            \quad \quad \texttt{top\_k=5} & 0.1293 & 0.0987 & 0.2129 & 0.1547 & 0.2211 & 0.0521 & 0.1447\\
            \quad GTR & & & & & & & \\
            \quad \quad \texttt{top\_k=1} & 0.4643 & 0.2291 & 0.7568 & 0.2366 & 0.4896 & 0.3118 & 0.4147 \\
            \quad \quad \texttt{top\_k=2} & 0.4397 & 0.2695 & 0.6035 & 0.2622 & 0.4632 & 0.3597 & 0.3996 \\
            \quad \quad \texttt{top\_k=3} & 0.3977 & 0.2714 & 0.4851 & 0.2584 & 0.4110 & 0.3260 & 0.3583 \\
            \quad \quad \texttt{top\_k=5} & 0.3234 & 0.2425 & 0.3466 & 0.2514 & 0.3285 & 0.2846 & 0.2962 \\
            \quad GTE-Qwen2 & & & & & & & \\
            \quad \quad \texttt{top\_k=1} & 0.4230 & 0.2015 & 0.7011 & 0.2188 & 0.4924 & 0.3435 & 0.3967 \\
            \quad \quad \texttt{top\_k=2} & 0.4100 & 0.2392 & 0.5841 & 0.2316 & 0.4751 & 0.3792 & 0.3865 \\
            \quad \quad \texttt{top\_k=3} & 0.3785 & 0.2442 & 0.4785 & 0.2415 & 0.4131 & 0.3314 & 0.3479 \\
            \quad \quad \texttt{top\_k=5} & 0.3133 & 0.2289 & 0.3467 & 0.2388 & 0.3174 & 0.2823 & 0.2879 \\
            \quad Qwen3 & & & & & & & \\
            \quad \quad \texttt{top\_k=1} & 0.5323 & 0.1574 & 0.7218 & 0.2810 & 0.4698 & 0.3151 & 0.4129 \\
            \quad \quad \texttt{top\_k=2} & 0.4963 & 0.1945 & 0.5950 & 0.2883 & 0.4592 & 0.3365 & 0.3950 \\
            \quad \quad \texttt{top\_k=3} & 0.4414 & 0.2057 & 0.4846 & 0.2892 & 0.4091 & 0.3205 & 0.3584 \\
            \quad \quad \texttt{top\_k=5} & 0.3546 & 0.1960 & 0.3496 & 0.2686 & 0.3186 & 0.2877 & 0.2959 \\
            \quad GritLM & & & & & & & \\
            \quad \quad \texttt{top\_k=1} & 0.5105 & 0.2722 & 0.8312 & 0.2890 & 0.5797 & 0.3650 & 0.4746 \\
            \quad \quad \texttt{top\_k=2} & 0.4762 & 0.3100 & 0.6473 & 0.3238 & 0.5351 & 0.3799 & 0.4454 \\
            \quad \quad \texttt{top\_k=3} & 0.4394 & 0.3008 & 0.5141 & 0.3258 & 0.4592 & 0.3398 & 0.3965 \\
            \quad \quad \texttt{top\_k=5} & 0.3573 & 0.2744 & 0.3616 & 0.2971 & 0.3445 & 0.3055 & 0.3234 \\
            \midrule
            \textbf{Autoregressive Models} & & & & & & & \\
            \quad SPScanner & & & & & & & \\
            \quad \quad \texttt{top\_k=1} & 0.4551 & 0.3232 & 0.7139 & 0.0656 & 0.3862 & 0.3878 & 0.3886 \\
            \quad \quad \texttt{top\_k=2} & 0.4732 & 0.3722 & 0.7210 & 0.1220 & 0.4268 & 0.4386 & 0.4256 \\
            \quad \quad \texttt{top\_k=3} & 0.4604 & 0.3712 & 0.6434 & 0.1354 & 0.4237 & 0.4188 & 0.4088 \\
            \quad \quad \texttt{top\_k=5} & 0.4060 & 0.3469 & 0.4965 & 0.1582 & 0.3893 & 0.3605 & 0.3596 \\
            \midrule
            \textbf{\name{} (Ours)} & & & & & & & \\
            \quad LLaMA-3.2 3B & & & & & & & \\
            \quad \quad \texttt{top\_k=1} & 0.5477 & \textbf{0.5030} & \underline{0.8550} & 0.2679 & \textbf{0.6065} & 0.3840 & 0.5274 \\
            \quad \quad \texttt{top\_k=2} & \underline{0.5704} & 0.4563 & 0.8394 & 0.3158 & \underline{0.6010} & \textbf{0.4843} & \underline{0.5445} \\
            \quad \quad \texttt{top\_k=3} & 0.5584 & 0.4618 & 0.8339 & \underline{0.3526} & 0.5998 & \underline{0.4738} & \textbf{0.5467} \\
            \quad \quad \texttt{top\_k=5} & 0.5133 & 0.4270 & 0.8009 & \textbf{0.3528} & 0.5810 & 0.4576 & 0.5221 \\
            \quad Qwen-2.5 3B & & & & & & & \\
            \quad \quad \texttt{top\_k=1} & 0.5583 & \underline{0.4827} & \textbf{0.8555} & 0.2220 & 0.5448 & 0.0891 & 0.4587 \\
            \quad \quad \texttt{top\_k=2} & \textbf{0.5774} & 0.4443 & 0.8454 & 0.2930 & 0.5709 & 0.3349 & 0.5110 \\
            \quad \quad \texttt{top\_k=3} & 0.5663 & 0.4551 & 0.8422 & 0.3106 & 0.5655 & 0.3215 & 0.5102 \\
            \quad \quad \texttt{top\_k=5} & 0.5155 & 0.4165 & 0.8093 & 0.3125 & 0.5468 & 0.3416 & 0.4904 \\
            \bottomrule
        \end{tabular}
    }
    \caption{Comparison of proposed method and baselines on single-document retrieval datasets with different choices of \texttt{top\_k} values, where best and second best results are marked in \textbf{bold} and \underline{underline}, respectively.}
    \label{table:retrieval-result-single-full}
\end{table*}

\section{Comparison of Efficiency on Single-document Retrieval Datasets}
\label{appendix:efficiency}

We compare the efficiency of \name{} with the baselines on long document retrieval datasets in Table \ref{table:retrieval-result-efficiency}. Although our method is considerable slower than sparse and small dense models, it is able to achieve similar latency as larger dense models including GTE, Qwen3, and GritLM, especially on datasets with larger average document lengths. 

\begin{table*}[!htb]
    \centering
    \tiny
    \resizebox{\textwidth}{!}{
        \begin{tabular}{l|ccccccc}
            \toprule
            & RepLiQA & ConditionalQA & NaturalQuestions & Qasper & QASA & LongBench-v2-Retrieval \\
            Average Length & 970.50 & 1298.13 & 2438.86 & 3442.26 & 4665.09 & 106025.49 \\
            \midrule
            BM25 & 0.0034 & 0.0049 & 0.0047 & 0.0054 & 0.0067 & 0.1005 \\
            DPR & 0.0585 & 0.1391 & 0.0824 & 0.1192 & 0.1207 & 3.4655 \\
            ANCE & 0.063 & 0.1412 & 0.1328 & 0.1764 & 0.2595 & 4.9592 \\
            CDE & 0.1305 & 0.3393 & 0.2442 & 0.3298 & 0.4222 & 10.1549 \\
            GTR & 0.1466 & 0.3607 & 0.2834 & 0.3913 & 0.5290 & 9.6679 \\
            GTE-Qwen2 & 0.4574 & 0.8310 & 1.1514 & 1.4872 & 2.7483 & 52.4382 \\
            Qwen3 & 0.5962 & 1.1245 & 1.4393 & 1.8504 & 3.3642 & 70.3171 \\
            GritLM & 0.6338 & 1.3311 & 1.8482 & 2.7332 & 3.5224 & 101.7907 \\
            SPScanner & 0.3790 & 0.7312 & 0.8978 & 1.1500 & 1.6098 & 46.4398 \\
            \name{}-LLaMA-3.2 3B & 0.9199 & 1.4303 & 1.5977 & 2.1085 & 2.9402 & 126.8405 \\
            \name{}-Qwen-2.5 3B & 0.9111 & 1.4051 & 1.6502 & 2.4614 & 2.8444 & 75.7912 \\
            \bottomrule
        \end{tabular}
    }
    \caption{Comparison of efficiency of proposed method and baselines on single-document retrieval datasets. The efficiency is measured in average processing time (both indexing and retrieval) in number of seconds for each sample.}
    \label{table:retrieval-result-efficiency}
\end{table*}

\section{Evaluation Results on Question Answering Datasets}
\label{appendix:qa}

The results on QA datasets are shown in Table \ref{table:qa-result}. \name{} achieves comparable performance with direct LLM generation with significantly less input tokens, and also outperforms SPScanner in RAG setting. 

\begin{table*}
    \centering
    \tiny
    \resizebox{\textwidth}{!}{
        \begin{tabular}{l|cc|cc|cc|cc|}
            \toprule
            & \multicolumn{2}{c|}{Qasper} & \multicolumn{2}{c|}{MultiFieldQA} & \multicolumn{2}{c|}{NarrativeQA} & \multicolumn{2}{c|}{Average} \\
            & F-1 & Avg Token Count & F-1 & Avg Token Count & F-1 & Avg Token Count & F-1 & Avg Token Count \\
            \midrule
            Llama-3.1 8B & & & & & & & & \\
            \quad Baseline & 0.3145 & 5026.46 & 0.5430 & 6996.87 & 0.2497 & 29883.02 & 0.3691 & 13968.78 \\
            \quad RAG with SPScanner & 0.2756 & 370.25 & 0.5173 & 449.22 & 0.1506 & 363.54 & 0.3145 & 394.34 \\
            \quad RAG with \name{}-Llama & 0.2929 & 392.975 & 0.5436 & 416.46 & 0.1654 & 322.09 & 0.3340 & 377.18 \\
            \quad RAG with \name{}-Qwen & 0.2697 & 382.84 & 0.5413 & 432.38 & 0.1408 & 350.92 & 0.3173 & 388.71 \\
            \midrule 
            Mistral-7B v0.3 & & & & & & & & \\
            \quad Baseline & 0.2933 & 5595.38 & 0.4942 & 7931.67 & 0.0724 & 35277.56 & 0.2866 & 16268.20 \\
            \quad RAG with SPScanner & 0.2485 & 375.04 & 0.4402 & 470.87 & 0.0997 & 359.56 & 0.2628 & 401.82 \\
            \quad RAG with \name{}-Llama & 0.2732 & 400.39 & 0.4854 & 431.09 & 0.1321 & 316.245 & 0.2969 & 382.57 \\
            \quad RAG with \name{}-Qwen & 0.2596 & 388.83 & 0.4904 & 451.57 & 0.1073 & 347.56 & 0.2858 & 395.98 \\
            \midrule
            Qwen-2.5 7B & & & & & & & & \\
            \quad Baseline & 0.3677 & 5108.42 & 0.5085 & 7205.38 & 0.1422 & 29920.98 & 0.3395 & 14078.26 \\
            \quad RAG with SPScanner & 0.2953 & 352.81 & 0.4766 & 438.25 & 0.1525 & 341.11 & 0.3081 & 377.39 \\
            \quad RAG with \name{}-Llama & 0.3220 & 376.13 & 0.5062 & 400.66 & 0.1746 & 299.74 & 0.3343 & 358.84 \\
            \quad RAG with \name{}-Qwen & 0.2927 & 365.56 & 0.5274 & 416.63 & 0.1532 & 328.52 & 0.3244 & 370.23 \\
            \midrule
            GPT-5 mini & & & & & & & & \\
            \quad Baseline & 0.3142 & 4974.21 & 0.4455 & 6899.97 & 0.2894 & 29520.40 & 0.3497 & 13798.19 \\
            \quad RAG with SPScanner & 0.2833 & 333.16 & 0.4460 & 409.96 & 0.2062 & 325.47 & 0.3118 & 356.20 \\
            \quad RAG with \name{}-Llama & 0.2980 & 355.96 & 0.4855 & 377.87 & 0.2393 & 284.10 & 0.3409 & 339.31 \\
            \quad RAG with \name{}-Qwen & 0.2971 & 346.15 & 0.4947 & 393.35 & 0.2076 & 312.40 & 0.3331 & 350.63 \\
            \bottomrule
        \end{tabular}
    }
    \caption{Comparison of proposed method and baselines on question answering tasks, where \textit{Baseline} represents using the full context to answer the question, while \textit{RAG} only uses the text chunks retrieved by the specified method to answer the question.}
    \label{table:qa-result}
\end{table*}

\section{Ablation Studies}

\begin{table*}[!htb]
    \centering
    \tiny
    \begin{tabular}{l|ccccccc|c}
        \toprule
        & QASA & Qasper & RepLiQA & ConditonalQA & NaturalQuestions & LongBench-v2-Retrieval & Average \\
        Average Length & 4665.09 & 3442.26 & 970.50 & 1298.13 & 2548.97 & 106025.49 & \\
        \midrule
        \name{} & 0.5584 & 0.4618 & \textbf{0.8339} & \textbf{0.3526} & \textbf{0.5998} & \textbf{0.4738} & \textbf{0.5467} \\
        Attention only for sentence scoring & 0.5344 & \textbf{0.4864} & 0.8296 & 0.3372 & 0.5912 & 0.3910 & 0.5283 \\
        Embedding only for sentence scoring & 0.4753 & 0.3164 & 0.7417 & 0.2598 & 0.4683 & 0.4242 & 0.4476 \\
        Removing entity graph & \textbf{0.5566} & 0.4468 & 0.7869 & 0.3371 & 0.5791 & 0.4443 & 0.5251 \\
        \bottomrule
    \end{tabular}
    \caption{Results of the ablation studies. The best performance for each dataset is marked with \textbf{bold}. The ablated settings generally have inferior performance compared to \name{}, demonstrating the effectiveness of the proposed approach.}
    \label{table:ablation}
\end{table*}

To verify the effectiveness of all components in the proposed retrieval model, we conducted ablation studies to measure the improvement brought by each component. We set up three variants of our system that removed attention scoring, embedding scoring, and entity graph, and evaluated the variants on the single-document retrieval datasets. The results are presented in Figure \ref{table:ablation}. In general, all components in our proposed pipeline brought substantial improvement, where attention scoring shows the most outstanding influence. This result is reasonable because attention scoring helps to model contextual and causal dependencies, two major challenges in long document retrieval. 

\section{Prompts}

We used the following prompt for all of our experiments: 

\begin{tcolorbox}
You are an AI assistant. User will you give you a task. Your goal is to complete the task as faithfully as you can.

Article: \{text\}

You are given an article and a question. Answer the question as concisely as you can, using a single phrase or sentence if possible.

Question: \{question\}

Answer: 
\end{tcolorbox}

\end{document}